\documentclass[a4paper,11pt]{article}
\pdfoutput=1
\usepackage{jheppub}
\usepackage{amsmath,amssymb,euscript}
\usepackage{braket}
\usepackage{slashed}
\usepackage{xspace}
\usepackage{color}
\usepackage{hyperref}
\usepackage{epsfig}
\usepackage{xcolor}
\usepackage{xspace}
\usepackage{verbatim}
\usepackage{multirow}
\usepackage{booktabs,graphicx}
\usepackage{mathtools}
\usepackage{tabulary}
\usepackage{soul}

\newcolumntype{K}[1]{>{\centering\arraybackslash}p{#1}}

\graphicspath{{figs/}}
\emergencystretch=\maxdimen
\renewcommand{\aa}{(aa)}
\newcommand{\ba}{(ba)}
\newcommand{\bb}{(bb)}
\newcommand{\aat}{(\widetilde{\phantom{aa}}\!\!\!\!\!\!aa)}
\newcommand{\bat}{(\widetilde{\phantom{aa}}\!\!\!\!\!\!ba)}
\newcommand{\bbt}{(\widetilde{\phantom{aa}}\!\!\!\!\!\!bb)}
\newcommand{\aab}{(\overline{aa})}
\newcommand{\bab}{(\overline{ba})}
\newcommand{\bbb}{(\overline{bb})}

\newcommand{\EOS}{\texttt{EOS}\xspace}
\newcommand{\wilson}{\texttt{wilson}\xspace}

\newcommand{\tev}{\, \mathrm{TeV}\xspace}
\newcommand{\gev}{\, \mathrm{GeV}\xspace}
\newcommand{\mev}{\, \mathrm{MeV}\xspace}

\makeatletter
\let\oldtheequation\theequation
\renewcommand\tagform@[1]{\maketag@@@{\ignorespaces#1\unskip\@@italiccorr}}
\renewcommand\theequation{(\oldtheequation)}
\makeatother

\begin{document}
\title{Resolving the Flavor Structure in the MFV-SMEFT}

\author[1,2]{Sebastian Bruggisser,}
\author[3,4]{Danny van Dyk,}
\author[1,5,6]{Susanne Westhoff\,}

\affiliation[1]{Institute for Theoretical Physics, Heidelberg University, 69120 Heidelberg, Germany}
\affiliation[2]{Department of Physics and Astronomy, Uppsala University, 75120 Uppsala, Sweden}
\affiliation[3]{Physik Department T31, Technische Universit\"at M\"unchen, 85748 Garching, Germany}
\affiliation[4]{Institute for Particle Physics Phenomenology, Durham University, Durham DH1 3LE, UK and
Department of Physics, Durham University, Durham DH1 3LE, UK}
\affiliation[5]{Institute for Mathematics, Astrophysics and Particle Physics, Radboud University, 6500 GL Nijmegen, The Netherlands}
\affiliation[6]{Nikhef, Science Park 105, 1098 XG Amsterdam, The Netherlands}

\emailAdd{bruggisser@thphys.uni-heidelberg.de}
\emailAdd{danny.van.dyk@gmail.com}
\emailAdd{susanne.westhoff@ru.nl}

		\begin{flushright}
		EOS-2022-04\\
		IPPP/22/62\\
		P3H-22-099\\
		TUM-HEP-1440/22\\
		NIKHEF-2022-016
		\end{flushright}

\abstract{%
We constrain the flavor structure of Wilson coefficients in the Standard Model Effective Field Theory (SMEFT) from data.
In the SMEFT, new physics effects in couplings of up-type and down-type quarks are related through the Cabibbo-Kobayashi-Maskawa mixing matrix. We exploit this relation to pin down potential new sources of flavor symmetry breaking in a global analysis of high- and low-energy data from the LHC, LEP, and $b$ factory experiments. We demonstrate the power of such an analysis by performing a combined fit of effective four-quark and two-quark couplings contributing to a large set of flavor, top-quark, electroweak, and dijet observables.
All four sectors are needed to fully resolve the flavor structure of the four-quark couplings without leaving blind directions in the parameter space.
Although we work in the framework of minimal flavor violation, our strategy applies as well to other flavor patterns, like $U(2)$ flavor symmetry or leptoquark scenarios.%
}

\maketitle
\flushbottom

\clearpage

\section{Introduction}\label{sec:intro}
\noindent The Standard Model Effective Field Theory (SMEFT)~\cite{Buchmuller:1985jz,Grzadkowski:2010es} is about to take on the role of a Standard Model for New Physics. Its purpose is to reveal or constrain patterns of subtle discrepancies with the Standard Model (SM) in several observables by probing virtual effects of heavy new physics in effective couplings of known particles. At the LHC, increasingly comprehensive and precise analyses of the parameter space of Wilson coefficients have been performed with top observables~\cite{Buckley:2015lku,Hartland:2019bjb,Brivio:2019ius}, with Higgs and electroweak observables~\cite{Biekoetter:2018ypq,Ellis:2018gqa,Falkowski:2019hvp}, as well as with combinations of these sectors~\cite{Ethier:2021bye}. Combined fits of high-energy and flavor observables are particularly powerful in resolving the SMEFT parameter space~\cite{Brod:2014hsa,Aebischer:2018iyb,Bissmann:2019gfc,Bissmann:2020mfi,Aoude:2020dwv,Bruggisser:2021duo}.

One of the biggest challenges in SMEFT analyses is the large number of effective couplings, which renders a global fit of the full parameter space computationally costly. The huge number of SMEFT parameters is mostly due to the unknown flavor structure of the underlying theory. In reality, however, the flavor structure of the Wilson coefficients is strongly constrained by the observed flavor hierarchies in mass and mixing among the SM fermions~\cite{Buchmuller:1985jz,Talbert:2021iqn}. In addition, in any concrete UV completion of the Standard Model, it seems likely that particle interactions follow a particular flavor pattern, as for instance in models addressing the origin of neutrino masses, the strong CP problem, or baryogenesis. Pinning down the flavor structure of the Wilson coefficients in SMEFT gives us insight into what the nature of such new interactions might be.

The flavor problem in SMEFT has been recently addressed by constructing flavor patterns for the Wilson coefficients that preserve the fermion masses and mixings in the Standard Model. Viable flavor patterns include Minimal Flavor Violation (MFV)~\cite{Buras:2000dm,DAmbrosio:2002vsn,Aoude:2020dwv,Bruggisser:2021duo}, $U(2)$ or $U(3)$ symmetries~\cite{Barbieri:2012uh,Faroughy:2020ina,Greljo:2022cah}, flavor alignment~\cite{Buchmuller:1985jz,Egana-Ugrinovic:2018znw}, and Froggatt-Nielsen scenarios including leptoquark couplings~\cite{Froggatt:1978nt,Bordone:2019uzc,Bordone:2020lnb,Talbert:2021iqn}. Assuming such an underlying pattern reduces the number of independent flavor parameters in the SMEFT and allows us to perform a global analysis of phenomenologically viable flavor structures in the first place.

Several analyses have investigated the leading SMEFT effects in the framework of specific flavor scenarios, see for instance Refs.~\cite{Efrati:2015eaa,Aoude:2020dwv,Ethier:2021bye}. This approach allows to identify differences between the various flavor scenarios. However, to \emph{resolve} the flavor structure of a given scenario, one has to disentangle flavor-conserving from flavor-breaking contributions to the Wilson coefficients. Within MFV, this has been demonstrated in Ref.~\cite{Bruggisser:2021duo} with a joint analysis of top-quark and flavor observables. In general, combined fits of observables involving up- and down-type quarks are a powerful tool to pin down the flavor structure of Wilson coefficients in the quark sector, because the effects are related through Cabibbo-Kobayashi-Maskawa (CKM) mixing. Similarly, the flavor structure of SMEFT coefficients in the lepton sector can be constrained by exploring correlated effects in processes with charged leptons and neutrinos.

In this work, we fully resolve the flavor structure of four-quark and two-quark SMEFT coefficients within the MFV framework. To this end, we perform a combined fit of top, flavor, $Z-$pole and dijet observables to data from the LHC, LEP, and $b$ factory experiments. We show how to disentangle flavor-conserving and flavor-breaking contributions and pin down the flavor structure of possible UV completions of the Standard Model.

This article is organized as follows. In \autoref{sec:smeft}, we parametrize the flavor structure of Wilson coefficients in MFV and introduce the framework for our analysis. In \autoref{sec:flavor}, we discuss SMEFT effects in $b-s$ transitions, reviewing the rare decays $B_s \to \mu^+\mu^-$ and $B\to X_s\gamma$ used previously in Ref.~\cite{Bruggisser:2021duo} and introducing $B_s-\bar{B}_s$ meson mixing. More details on the flavor observables can be found in \autoref{app:expressions}. In \autoref{sec:weak-scale}, we consider observables at the weak scale. We review the top observables used in Ref.~\cite{Bruggisser:2021duo} and analyze in addition the flavor structure in $t\bar{t}b\bar{b}$ production and in $Z-$pole observables. In \autoref{sec:dijets}, we investigate dijet angular distributions at the LHC, which probe SMEFT contributions at the TeV scale. The results of our global fit are presented in \autoref{sec:global}. For the first time, we constrain the full flavor structure of effective four-quark couplings in MFV. We conclude in \autoref{sec:conclusions}.

\section{Flavor in the SMEFT}\label{sec:smeft}
\noindent
Our starting point is the general effective Lagrangian in the SMEFT,
\begin{align}\label{eq:smeft-lagrangian}
\mathcal{L}_{\rm SMEFT} = \sum_a \frac{C_a}{\Lambda^2}\,O_a + \dots
\end{align}
It describes low-energy effects of potential new physics above a cutoff scale $\Lambda$ in terms of local operators $O_a$ and their Wilson coefficients $C_a$. The sum runs over all dimension-six operators that respect the SM gauge symmetries. The dots stand for higher-dimensional operators, which we do not consider in this work. Hermiticity of the SMEFT Lagrangian is implied and leads to relations between the Wilson coefficients as discussed below.

We focus on four-quark operators with only left-handed quark fields, which feature a particularly rich flavor structure. A similar analysis could be conducted for four-quark operators with only right-handed quarks or mixed chiralities. Since four-quark operators mix with two-quark operators under the renormalization group (RG), we consider a set of four-quark and two-quark operators that is closed under the RG evolution to leading-logarithmic (LL) accuracy~\cite{Alonso:2013hga}.
In the Warsaw basis~\cite{Grzadkowski:2010es}, these operators are defined as
\begin{gather}\label{eq:operators:4q}
\begin{aligned}
O_{qq}^{(1),klmn} & = (\overline{Q}^k\gamma^\mu Q^l)(\overline{Q}^m\gamma_\mu Q^n) \\
O_{qq}^{(3),klmn} & = (\overline{Q}^k\gamma^\mu \tau^I Q^l)(\overline{Q}^m\gamma_\mu \tau^I Q^n)
\end{aligned}\\
\label{eq:operators:2q}
\begin{aligned}
O_{\phi q}^{(1),kl} & = (\phi^\dagger \stackrel{\longleftrightarrow}{iD_\mu} \phi)(\overline{Q}^k\gamma^\mu Q^l)\\
O_{\phi q}^{(3),kl} & = (\phi^\dagger \stackrel{\longleftrightarrow}{iD_\mu^I} \phi)(\overline{Q}^k\gamma^\mu\tau^I Q^l)\,.
\end{aligned}
\end{gather}
Here $Q$ is a weak doublet of left-handed quarks; $\phi$ is the Higgs doublet; and $\tau^I$ are the generators of weak interactions. The indices $\{k,l,m,n\} \in \{1,2,3\}$ denote the three quark generations.
In what follows, we use $w=\{1,3\}$ to denote operators $O^{(w)}$ or their Wilson coefficients $C^{(w)}$ with either a weak gauge singlet or triplet structure.

The Wilson coefficients $C_{\phi q}^{(w)}$ are $3\times 3$ matrices and $C_{qq}^{(w)}$ are $3\times 3 \times 3 \times 3$ tensors in flavor space. For the two-quark coefficients, hermiticity of the SMEFT Lagrangian requires that
\begin{align}
    (C_{\phi q}^{(w)})_{kl} = (C_{\phi q}^{(w)})_{lk}^\ast\,.
\end{align}
The number of independent real parameters per operator is thus $9$. For four-quark couplings with identical quark fields, the following relations hold~\cite{Aguilar-Saavedra:2018ksv}
\begin{align}
(C_{qq}^{(w)})_{klmn} & = (C_{qq}^{(w)})_{lknm}^\ast\qquad \text{and}\qquad
(C_{qq}^{(w)})_{klmn} = (C_{qq}^{(w)})_{mnkl}\,.
\end{align}
The number of independent parameters per operator is thus $27$~\cite{Fuentes-Martin:2020zaz}. In total, the Wilson coefficients for the operator set from \autoref{eq:operators:4q} and \autoref{eq:operators:2q} introduce 72 independent real parameters.

Besides these theory constraints, some of the directions in flavor space are strongly constrained by the observation of quark mass hierarchies and CKM mixing. To ensure that this pattern is reflected in new particle interactions, we apply the principle of Minimal Flavor Violation to the flavor structure of the Wilson coefficients~\cite{Buras:2000dm,DAmbrosio:2002vsn}.

All gauge interactions of quarks in the Standard Model respect the flavor symmetry
\begin{align}
\mathcal{G}_F = U(3)_Q \times U(3)_U\times U(3)_D\,.
\end{align}
This symmetry is broken by the Yukawa couplings, which transform under $\mathcal{G}_F$ as
\begin{align}
Y_U:\ (3,\overline{3},1)\,,\qquad Y_D:\ (3,1,\overline{3})\,.
\end{align}
The MFV framework is based on the idea that the Yukawa couplings are the only sources of flavor symmetry breaking, in the Standard Model and also in extensions. Under the assumption of MFV, we can describe the flavor structure of left-handed quark currents $\overline{Q}\dots Q$ in terms of a $3\times 3$ matrix, transforming under $\mathcal{G}_F$ as
\begin{align}
\mathcal{A}_Q:\ (3\times \overline{3},1,1)\,.
\end{align}
Expanding in terms of the Yukawa matrices, we obtain
\begin{align}\label{eq:mfv-expansion}
\mathcal{A}_Q & = a\, {\bf 1} + b\,Y_U Y_U^\dagger + c\, Y_D Y_D^\dagger + \dots
\end{align}
where $a,b,c$ are in general complex parameters. Here we have kept only the leading terms in $Y_U, Y_D$. Higher orders in $Y_U$ up to the fourth power are included in our analysis; higher orders in $Y_D$ are neglected, because they are suppressed by the small down-quark Yukawa couplings.

For two-quark operators, we obtain the flavor structure in MFV directly from $\mathcal{A}_Q$,
\begin{align}
 (\mathcal{A}_Q)_{kl}\, (\overline{Q}^k\gamma_\mu\,Q^l)\,.
\end{align}
For four-quark operators, the flavor structure reads
\begin{align}\label{eq:structure-4q}
\big[ (\mathcal{A}_Q)_{kl} (\mathcal{A}_Q)_{mn} + (\widetilde{\mathcal{A}}_Q)_{kn}(\widetilde{\mathcal{A}}_Q)_{ml} \big]\, (\overline{Q}^k\gamma^\mu\,Q^l)(\overline{Q}^m\gamma_\mu\,Q^n)\,,
\end{align}
where $\mathcal{A}$ and $\widetilde{\mathcal{A}}$ refer to two possible flavor contractions $(kl)(mn)$ and $(kn)(ml)$.

Without losing generality, we work in the Warsaw \emph{up mass basis}\footnote{%
    see e.g.~the definition as part of the \texttt{WCxf} software~\cite{Aebischer:2017ugx}.%
}, where the gauge eigenstates of left-handed quarks are aligned with the mass eigenstates of up-type quarks,
\begin{align}\label{eq:up-alignment}
Q^k = \begin{pmatrix} u_L^k \\V_{kl} d^l_L \end{pmatrix},
\end{align}
where $V$ is the CKM matrix and $u_L,d_L$ are the mass eigenstates of (left-handed) up- and down-type quarks. In the up mass basis, the Yukawa matrices are 
\begin{align}
 Y_U = Y_u\,, \quad Y_D = V Y_d\,,
\end{align}
with the physical Yukawa couplings $Y_u = \text{diag}(y_u,y_c,y_t)$, $Y_d = \text{diag}(y_d,y_s,y_b)$.
In this framework, we write the Wilson coefficients for the two-quark operators in \autoref{eq:operators:2q} as
\begin{equation}
    \label{eq:wc-mfv:2q}
    \begin{aligned}
    (C_{\phi q}^{(w)})_{kk} & = a^{(w)}  + b^{(w)} y_t^2\,\delta_{k3}\,.\\
    \end{aligned}
\end{equation}
For the four-quark operators in \autoref{eq:operators:4q}, we parametrize the flavor structure as
\begin{equation}
    \label{eq:wc-mfv:4q}
    \begin{aligned}
    (C_{qq}^{(w)})_{kkii} & = \aa^{(w)} + \ba^{(w)} y_t^2\, \delta_{k3}\\
    (C_{qq}^{(w)})_{kiik} & = \aat^{(w)} + \bat^{(w)} y_t^2\, \delta_{k3}\\
    (C_{qq}^{(w)})_{3333} & = \aab^{(w)} + 2\,\bab^{(w)} y_t^2 + \bbb^{(w)} y_t^4\,,
    \end{aligned}
\end{equation}
where $i=\{1,2\}$ labels quarks from the first and second generation. Coefficients with other flavor indices are zero. We have introduced the combinations
\begin{align}
      \aab^{(w)} = \aa^{(w)} + \aat^{(w)},\quad \bab^{(w)} = \ba^{(w)} + \bat^{(w)},
\end{align}
where parameters with and without a tilde originate from $\widetilde{\mathcal{A}}_Q\widetilde{\mathcal{A}}_Q$ and $\mathcal{A}_Q\mathcal{A}_Q$, see \autoref{eq:structure-4q}. For general flavor structures, there are also two possible flavor contractions $\bb^{(w)}$ and $\bbt^{(w)}$. In MFV, however, they can only be probed in the combination
\begin{align}
\bbb^{(w)} = \bb^{(w)} + \bbt^{(w)}.
\end{align}
We call this setup the \emph{MFV-SMEFT} and refer to its parameters
as \emph{flavor parameters}.\footnote{For earlier approaches to the MFV-SMEFT, see Refs.~\cite{Aoude:2020dwv,Faroughy:2020ina}.} The MFV assumption implies that the SMEFT coefficients $C$ and the flavor parameters $a,b,\dots,\bbb$ are real-valued. For the Wilson coefficients in \autoref{eq:wc-mfv:2q} and \autoref{eq:wc-mfv:4q}, the flavor structure is thus described by $2\times 2$ real parameters for the two-quark operators
\begin{align}\label{eq:2q-flavor}
    \big\{a^{(w)},\ b^{(w)}\big\}
\end{align}
and $2\times 5$ real parameters for the four-quark operators
\begin{align}\label{eq:4q-flavor}
    \big\{\aa^{(w)},\ \aat^{(w)},\ \ba^{(w)},\ \bat^{(w)},\  \bbb^{(w)}\big\}\,.
\end{align}
In total, in the MFV-SMEFT the set of operators in \autoref{eq:operators:4q} and \autoref{eq:operators:2q}
is described by $14$ flavor parameters. Among them, the parameters $a^{(w)}$, $\aa^{(w)}$ and $\aat^{(w)}$ denote flavor-universal contributions.
Flavor breaking in one quark bilinear is encoded in $b^{(w)}$, $\ba^{(w)}$ and $\bat^{(w)}$;
flavor breaking in both bilinears is parametrized by $\bbb^{(w)}$.
Here and throughout our analysis, we only consider flavor-breaking terms from the top Yukawa coupling and neglect subleading contributions of $\mathcal{O}(y_b^2)$. 
For later convenience, we introduce the combinations
\begin{align}
    a^{(+)} = a^{(1)} + a^{(3)},\qquad a^{(-)} & = a^{(1)} - a^{(3)},
\end{align}
and analogously for all remaining flavor parameters. At tree level, the combinations $(-)$ and $(+)$ are probed in weak neutral currents with two up-type quarks and down-type quarks, respectively. Beyond tree level, small corrections to this assignment occur.

In order to probe this 14-dimensional parameter space, observables from different sectors in particle physics need to be combined. For example, several flavor and LHC observables are sensitive to the four-quark operators from \autoref{eq:operators:4q}, but involve quarks with different charges under the gauge and flavor groups. Electroweak observables are very sensitive to the two-quark operators from \autoref{eq:operators:2q}. We resolve the full flavor space by combining a variety of observables at different energy scales in a global analysis: flavor observables at the GeV scale, $Z-$pole and top observables around the weak scale, and dijet production observables at the TeV scale. This combination requires a consistent treatment of the RG evolution and mixing of operators across the different scales.\footnote{%
    The RG evolution of the flavor structure of SMEFT coefficients has been investigated more generally in Ref.~\cite{Machado:2022ozb}.
}
We report the bounds on the flavor parameters from \autoref{eq:2q-flavor} and \autoref{eq:4q-flavor} defined at the scale $\mu_0 = 2.4\,$TeV in units of TeV$^2/\Lambda^2$. This common reference scale $\mu_0$ is motivated by the observables that probe the highest energies, namely angular distributions of dijets at the LHC (see \autoref{sec:dijets}).\footnote{%
    Notice that this scale choice differs from Ref.~\cite{Bruggisser:2021duo}, where $\mu_0 = m_t$ was used.
}

To prepare for the global analysis, we analyze the SMEFT effects in selected observables from each sector and determine which directions in flavor space they probe.

\section{GeV scale: flavor observables}\label{sec:flavor}
To describe flavor observables in our analysis, we match the relevant SMEFT amplitudes onto the Weak Effective Theory (WET). The flavor observables are expressed in terms of WET coefficients, which are linear combinations of SMEFT coefficients. We combine the predictions and measurements of several flavor observables in a likelihood function, which can be evaluated at any point in the parameter space of the SMEFT coefficients from \autoref{eq:operators:4q} and
\eqref{eq:operators:2q}. Our treatment of the flavor observables and the RG evolution of the involved Wilson coefficients closely follows the setup of Ref.~\cite{Bruggisser:2021duo}.

The WET is commonly split into so-called \emph{sectors} of operators, which are distinguished by the flavor quantum numbers of the involved fields~\cite{Aebischer:2017gaw,Jenkins:2017jig}.
For this analysis, we use the sectors $sb$, $sb\mu\mu$, and $sbsb$.
The effective Lagrangian for each sector takes the form
\begin{align}\label{eq:wet-lag}
    \mathcal{L}_{\mathcal{S}} = \sum_\alpha \mathcal{C}^{\mathcal{S}}_\alpha\,\mathcal{O}^{\mathcal{S}}_\alpha + h.c. + \dots\,,
\end{align}
where $\mathcal{O}^{\mathcal{S}}_\alpha$ are dimension-six WET operators in the sector $\mathcal{S} = \{sb,sb\mu\mu,sbsb\}$,
and the dots indicate operators of mass dimension larger than six. The Wilson coefficients $\mathcal{C}^{\mathcal{S}}_\alpha$ contain both SM and SMEFT contributions.

To obtain the SM contribution to the WET coefficients, we match the full amplitude in the electroweak theory onto the WET amplitude at the scale $M_Z$. We then RG-evolve the WET coefficients to a low-energy scale of a few GeV, $\mu_{\mathcal{S}}$.
For the RG evolution of the SM coefficients in the WET, we use at least next-to-leading logarithmic (NLL) accuracy.
Next-to-next-to-leading logarithmic (NNLL) accuracy is used if available.

To obtain the SMEFT contribution to the WET coefficients, we follow a two-step procedure. First, we perform the SMEFT-to-WET matching at the scale $M_Z$, using matching relations at the one-loop level~\cite{Dekens:2019ept}. For SMEFT contributions of $\mathcal{O}(\Lambda^{-2})$, the matching conditions are linear in the SMEFT coefficients. We do not consider SMEFT contributions to WET coefficients beyond $\mathcal{O}(\Lambda^{-2})$, which in general would require including dimension-eight SMEFT operators to renormalize.
 In a second step, we express the SMEFT coefficients at the scale $M_Z$ in terms of SMEFT coefficients at the higher scale $\mu_0 = 2.4\,$TeV. The RG evolution between $M_Z$ and $\mu_0$ is done at leading-logarithmic (LL) accuracy~\cite{Aebischer:2017gaw}. As discussed in Ref.~\cite{Bruggisser:2021duo}, this procedure can and should be improved to NLL accuracy as soon as the two-loop anomalous dimensions for the full basis of WET operators become available. This would be particularly relevant for some observables in our analysis, for which we find a strong dependence on the renormalization scale, see \autoref{sec:sbsb}.

With this procedure, we obtain the WET coefficients at the low scale $\mu_{\mathcal{S}}$ as a \emph{linear combination} of the SMEFT coefficients at the high scale $\mu_0$,
\begin{align}\label{eq:matching}
    \mathcal{C}_\alpha^{\mathcal{S}}(\mu_{\mathcal{S}}) = \mathcal{C}_{\alpha,{\rm SM}}^{\mathcal{S}}(\mu_{\mathcal{S}}) + \sum_{a} M_{\alpha a}(\mu_\mathcal{S},\mu_0)\,C_a(\mu_0),
\end{align}
where $\mathcal{C}_{\alpha,{\rm SM}}^{\mathcal{S}}$ is the SM contribution to the WET coefficient and the factors $M_{\alpha a}$ encode the running and matching. The SMEFT coefficients $C_a$ are linear combinations of the flavor parameters from \autoref{eq:2q-flavor} and \autoref{eq:4q-flavor}, defined at $\mu_0 = 2.4\,$TeV. The index $a$ runs over all relevant SMEFT operators. For the low scale, we use $\mu_{\mathcal{S}} = 4.2\,$GeV in all three flavor sectors.

To leading order in the WET, the decay rates $\Gamma$ of flavor-changing processes are sesquilinear polynomials of the dimension-six WET coefficients from \autoref{eq:matching},
\begin{align}\label{eq:decay-rate-wet}
    \Gamma = \sum_{\alpha,\beta} \mathcal{C}_\alpha \mathcal{C}_\beta^\ast\,\Gamma^{\alpha\beta},
\end{align}
where $\Gamma^{\alpha\beta}$ are the corresponding contributions to the observable. Contributions of dimension-four interactions are absent here; they only occur in flavor-conserving processes. As a consequence, no interference of such amplitudes with dimension-eight operators is possible, so that dimension-eight operators do not contribute in \autoref{eq:decay-rate-wet}.

Inserting the WET coefficients from \autoref{eq:matching} into \autoref{eq:decay-rate-wet}, we finally obtain the decay rates as sesquilinear polynomials in the dimension-six SMEFT coefficients,
\begin{align}\label{eq:decay-rate-smeft}
    \Gamma = \Gamma_{\rm SM} + \sum_{a}\frac{C_a}{\Lambda^2}\,\Gamma_{\rm int}^a + \sum_{a,b}\frac{C_a C_b}{\Lambda^4}\,\Gamma_{\rm SMEFT}^{ab}.
\end{align}
Here $\Gamma_{\rm SM}$ denotes the decay rate in the Standard Model, and $\Gamma_{\rm int}^a$, $\Gamma_{\rm SMEFT}^{ab}$ are the (normalized) contributions from SM-SMEFT and SMEFT-SMEFT operator interference, respectively. For meson decays, the measurements of partial rates $\Gamma_i$ are typically reported as branching ratios $\mathcal{B}_i = \Gamma_i/\Gamma_{\rm tot}$. Since an accurate prediction of the total decay rate $\Gamma_{\rm tot}$ is impossible due to large hadronic uncertainties, we follow the common procedure to use the measurement of $\Gamma_{\rm tot}$ to predict $\mathcal{B}_i$. The flavor structure of the branching ratios in SMEFT is thus the same as for the partial decay rates.

While some SMEFT analyses of high-energy observables choose to truncate the power expansion of the observables at $\mathcal{O}(\Lambda^{-2})$, this is not an option for the meson decays. Removing the quadratic contributions in the SMEFT coefficients in \autoref{eq:decay-rate-smeft} would require to treat the SM and SMEFT contributions to the WET coefficients in \autoref{eq:matching} differently. This is inconsistent with the idea of WET as an effective field theory for flavor observables, whose structure is agnostic of UV completions above the weak scale.

On the technical side, for the RG evolution of the Wilson coefficients in SMEFT and WET we use the \wilson software~\cite{Aebischer:2018bkb}.
The predictions of flavor observables in terms of WET coefficients are obtained using the \EOS software~\cite{vanDyk:2021sup}.
These predictions involve hadronic matrix elements that introduce hadronic parametric uncertainties.
Unless otherwise stated, we treat these uncertainties as in Ref.~\cite{Bruggisser:2021duo}.
 
In what follows, we provide details on each of the WET sectors. In particular, we discuss the relevant operators and their flavor structure, the implementation of the flavor observables in our analysis, and the likelihood of flavor observables used in the fit.

\subsection{The $sb$ sector}\label{sec:sb}
The $sb$ sector is comprised of four WET operators with a dipole structure: the electromagnetic dipole operator
$\mathcal{O}^{sb}_{7}$, its chiral counterpart $\mathcal{O}^{sb}_{7'}$, the chromomagnetic dipole operator
$\mathcal{O}^{sb}_{8}$, and its chiral counterpart $\mathcal{O}^{sb}_{8'}$. In our convention they read
\begin{equation}
\label{eq:wet-sb}
\begin{aligned}
    \mathcal{O}^{sb}_{7}
        & = \frac{4G_F }{\sqrt{2}} V_{tb}V_{ts}^\ast
        \frac{e}{16\pi^2}m_b\left(\overline{s}\, \sigma^{\mu\nu} P_{R} b\right)F_{\mu\nu}\,, \\
    \mathcal{O}^{sb}_{7'}
        & = \frac{4G_F }{\sqrt{2}} V_{tb}V_{ts}^\ast
        \frac{e}{16\pi^2}m_b\left(\overline{s}\, \sigma^{\mu\nu} P_{L} b\right)F_{\mu\nu}\,, \\
    \mathcal{O}^{sb}_{8}
        & = \frac{4G_F}{\sqrt{2}}V_{tb}V_{ts}^\ast
        \frac{g_s}{16\pi^2}m_b\left(\overline{s}\, \sigma^{\mu\nu}T^A P_{R} b\right)G_{\mu\nu}^A\,, \\
    \mathcal{O}^{sb}_{8'}
        & = \frac{4G_F}{\sqrt{2}}V_{tb}V_{ts}^\ast
        \frac{g_s}{16\pi^2}m_b\left(\overline{s}\, \sigma^{\mu\nu}T^A P_{L} b\right)G_{\mu\nu}^A\,.
\end{aligned}
\end{equation}
Here, $e$ and $g_s$ are the electromagnetic and QCD coupling constants,
$F_{\mu\nu}$ and $G_{\mu\nu}^A$ are the field strength tensors, $T^A$ are the $SU(3)_C$ generators, $V_{ij}$ are the CKM matrix elements,
and $P_{L},P_R$ are chiral projectors. For the Fermi constant $G_F$, we use the value extracted from muon decay.
The Wilson coefficients $\mathcal{C}^{sb}_\alpha$ are renormalized in the $\overline{\text{MS}}$ scheme at the scale $\mu_{\mathcal{S}} = 4.2\gev$.

The inclusive decay $\bar{B}\to X_s\gamma$ is sensitive to these four operators. The world average for measurements of its branching fraction is~\cite{%
CLEO:2001gsa,BaBar:2007yhb,Belle:2009nth,BaBar:2012fqh,BaBar:2012eja,%
Belle:2014nmp,HFLAV:2019otj,ParticleDataGroup:2020ssz}
\begin{equation}
    \mathcal{B}(\bar{B}\to X_s\gamma)\big|_{E_\gamma \geq 1.9\gev}
        = (3.49 \pm 0.19) \times 10^{-4}\,.
\end{equation}
The branching ratio $\mathcal{B}(\bar{B}\to X_s\gamma)$ is particularly sensitive to $\mathcal{C}^{sb}_{7}$ and $\mathcal{C}^{sb}_{7'}$, with numerically sub-leading contributions of $\mathcal{C}^{sb}_{8}$ and $\mathcal{C}^{sb}_{8'}$. The matching procedure and RG evolution translates this into a sensitivity to the SMEFT parameters.
In the MFV-SMEFT, the branching ratio is a sesquilinear polynomial in the flavor parameters, following the structure of \autoref{eq:decay-rate-smeft}.
We find a dominant sensitivity to the interference between the SM contribution and the parameters
$b^{(+)}$ and $a^{(3)}$, with a relative suppression compared to the SM term by roughly
one order of magnitude. Somewhat further suppressed enter the parameters $\aat^{(3)}$, $\bat^{(3)}$,
and $\bbb^{(3)}$.
In \autoref{tab:sb:BstoXsgamma} in the appendix, we give the numerically leading contributions to this polynomial.\footnote{The numerical results differ from the predictions in Ref.~\cite{Bruggisser:2021duo}, due to the change of the scale $\mu_0$ and due to including the full contributions of four-quark operators.}

\subsection{The $sb\mu\mu$ sector}\label{sec:sbmumu}
In our analysis, we include two operators from the $sb\mu\mu$ sector of the WET,
\begin{equation}
\label{eq:wet-sbmumu}
\begin{aligned}
    \mathcal{O}^{sb\mu\mu}_{10}
        & = \frac{4G_F}{\sqrt{2}} V_{tb}V_{ts}^\ast
            \frac{e^2}{16\pi^2}\left(\overline{s}\,\gamma_\mu P_{L(R)} b\right)\left(\overline{\mu}\gamma^\mu\gamma_5 \mu\right)\,,\\
    \mathcal{O}^{sb\mu\mu}_{10'}
        & = \frac{4G_F}{\sqrt{2}} V_{tb}V_{ts}^\ast
            \frac{e^2}{16\pi^2}\left(\overline{s}\,\gamma_\mu P_{L(R)} b\right)\left(\overline{\mu}\gamma^\mu\gamma_5 \mu\right)\,.
\end{aligned}
\end{equation}
We constrain the Wilson coefficients $\mathcal{C}_{10}(\mu_{\mathcal{S}}),\mathcal{C}_{10'}(\mu_{\mathcal{S}})$ from measurements of the branching fraction for $\bar{B}_s\to \mu^+\mu^-$ decays.
The world average of measurements~\cite{LHCb:2017rmj,ATLAS:2018cur,CMS:2019bbr,LHCb:2020zud} is given by\footnote{%
    As in Ref.~\cite{Bruggisser:2021duo}, we symmetrize the uncertainties around the central value by using the larger uncertainty in both directions.
    This is required for compatibility with the treatment of uncertainties in the \texttt{sfitter}
    software~\cite{Brivio:2022hrb}.
}
\begin{equation}
    \mathcal{B}(\bar{B}_s\to \mu^+\mu^-) = \left(2.69^{+0.37}_{-0.35}\right) \times 10^{-9}\,.
\end{equation}
Other WET coefficients within the $sb\mu\mu$ sector are not relevant to our discussion,
since they either do not contribute to the prediction of $\mathcal{B}(\bar{B}_s\to \mu^+\mu^-)$\footnote{%
    In particular, $\mathcal{C}_9$ and four-quark operators do not contribute to
    $\mathcal{B}(\bar{B}_s\to \mu^+\mu^-)$ at leading order in $\alpha_e$.
}
or do not receive contributions of our set of MFV-SMEFT parameters \autoref{eq:wc-mfv:2q} and \autoref{eq:wc-mfv:4q} in the SMEFT-to-WET matching.\\
\indent The theory expression for the $\bar{B}_s\to \mu^+\mu^-$ branching ratio is a sesquilinear polynomial
in the flavor parameters, see \autoref{tab:sb:Bstomumu} in the appendix. The sensitivity to $b^{(+)}$ is particularly strong: Quadratic and linear contributions of $b^{(+)}$ are enhanced by factors of about $30$ and $10$ with respect to the SM contribution, respectively. Interference of $b^{(+)}$ with contributions of $a^{(3)}$ and various four-quark parameters are also about ten-fold enhanced compared to the Standard Model.

\subsection{The $sbsb$ sector}\label{sec:sbsb}
The $sbsb$ sector of the WET is crucial to describe $B_s$--$\overline{B}_s$ meson mixing. Since $B_s$--$\overline{B}_s$ mixing was not included in our previous analysis~\cite{Bruggisser:2021duo}, we discuss this sector in more detail. For $sbsb$ interactions, we normalize the WET Lagrangian as
\begin{equation}
    \mathcal{L}_{sbsb} = \frac{4 G_F}{\sqrt{2}} \, \left(V_{tb} V_{ts}^{\ast}\right)^2 \sum_\alpha \mathcal{C}^{sbsb}_\alpha \mathcal{O}^{sbsb}_\alpha + h.c.\,.
\end{equation}
We define the operators $\mathcal{O}^{sbsb}_\alpha$ by closely following the so-called Bern basis,\footnote{%
    Compared to the Bern basis~\cite{Aebischer:2017gaw}, we have factored out the CKM matrix elements to be consistent
    with the notation used for the other sectors and with the normalisation used in the \EOS software.}
\begin{align}\label{eq:sbsb-operators}
    \mathcal{O}_{1^{(\prime)}}^{sbsb} & = \left[\bar{q} \gamma_\mu P_{L(R)} b      \right]\,\left[\bar{q}\gamma^\mu P_{L(R)} b\right],        &
    \mathcal{O}_{2^{(\prime)}}^{sbsb} & = \left[\bar{q}            P_{L(R)} b      \right]\,\left[\bar{q}           P_{L(R)} b\right],        \\\nonumber
    \mathcal{O}_{3^{(\prime)}}^{sbsb} & = \left[\bar{q}_r          P_{L(R)} b_s\right]\,\left[\bar{q}_s     P_{L(R)} b_r\right], &
    \mathcal{O}_{4}^{sbsb}            & = \left[\bar{q}            P_{L}    b      \right]\,\left[\bar{q}           P_{R}    b\right],        \\\nonumber
    \mathcal{O}_{5}^{sbsb}            & = \left[\bar{q}_r          P_{L}    b_s\right]\,\left[\bar{q}_s     P_{R}    b_r\right],
\end{align}
where $\{r,s\}$ are color indices in the fundamental representation of $SU(3)_C$.

To constrain the WET coefficients, we use the difference of mass eigenstates in the $B_s$--$\overline{B}_s$ system,
$\Delta m_s$. The PDG world average of measurements of $\Delta m_s$
is~\cite{CDF:2006imy,LHCb:2011vae,LHCb:2013fep,LHCb:2013lrq,LHCb:2019nin,CMS:2020efq,LHCb:2020qag,LHCb:2021moh,ParticleDataGroup:2020ssz}
\begin{equation}
    \Delta m_s = 17.746 \pm 0.029\, \text{ps}^{-1}\,.
\end{equation}
The prediction of $\Delta m_s$ involves a number of hadronic matrix elements~\cite{Dowdall:2019bea}
\begin{equation}
    R_s^{(i)} \equiv \frac{4 \bra{B_s} \mathcal{O}^{sbsb}_i\ket{\bar{B}_s}}{\left( f_{B_s} M_{B_s}\right)^2}\,,
\end{equation}
which are normalized to the $B_s$ decay constant, $f_{B_s} = (230.7 \pm 1.3)\,\mev$~\cite{Bazavov:2017lyh}.
For our analysis, we only need the matrix element $R_s^{(1)}$ that multiplies $\mathcal{C}^{sbsb}_1$,
for which we use the lattice QCD results from Ref.~\cite{Dowdall:2019bea}. We do not account for the correlation
between $f_{B_s}$ and $R_s^{(1)}$, since
the uncertainty on $R_s^{(1)}$ induced by $f_{B_s}$ is negligible compared to the (still) sizeable uncertainties on the hadronic matrix elements. The decay constant $f_{B_s}$ also enters the prediction of $\mathcal{B}(\bar{B}_s \to \mu^+\mu^-)$, where we include the uncertainty on $f_{B_s}$ as the only hadronic uncertainty.

In the Standard Model, the only Wilson coefficient that contributes to $\Delta m_s$ is~\cite{Buras:2001ra}
\begin{align}
    \mathcal{C}^{sbsb}_{1,\text{SM}}(4.2\,\text{GeV}) = 1.31\times 10^{-3}.
\end{align}
In the MFV-SMEFT, the operators from \autoref{eq:operators:4q} and \autoref{eq:operators:2q} contribute to $\mathcal{C}^{sbsb}_1$
through tree-level SMEFT-to-WET matching and through the RG evolution within the SMEFT.\footnote{%
    For a comprehensive general analysis of SMEFT contributions to meson mixing, see Refs.~\cite{Endo:2018gdn,Aebischer:2020dsw}.}
Contributions to all other WET operators from \autoref{eq:sbsb-operators} are suppressed by powers of $y_b$ or $y_s$ and will not be considered here.

\begin{figure}[t]
    \centering
    \includegraphics[width=\textwidth]{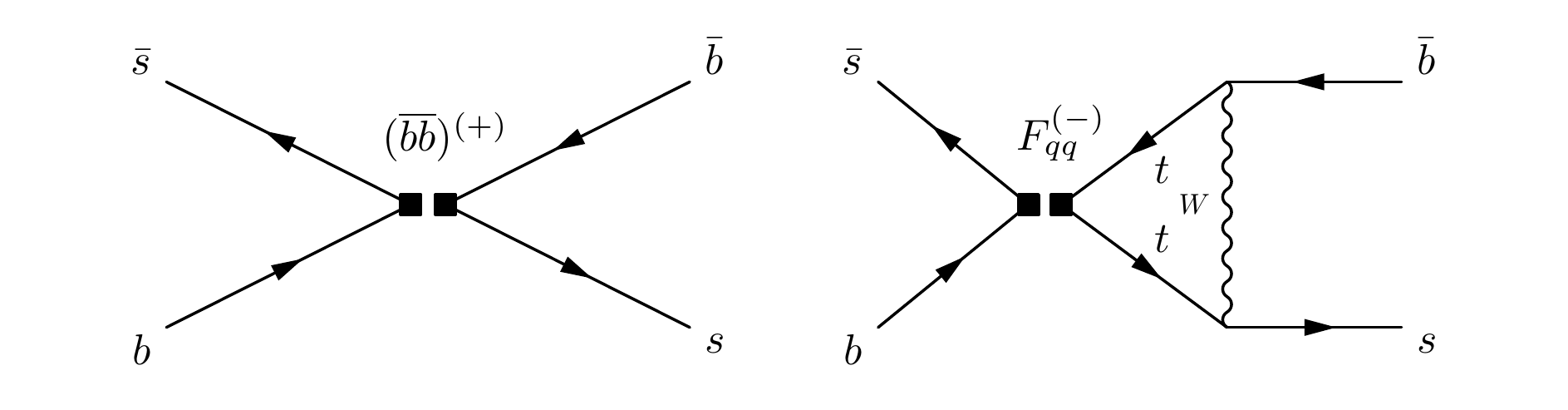}
    \caption{%
        Illustration of SMEFT operator contributions to $B_s$ mixing observables. Left: tree-level topology, with insertion of $\bbb^{(+)}$. Right: one-loop topology, involving a charged current with the
        insertion of $F_{qq}^{(-)}$; see \autoref{eq:sbsb-loop}.
        The operator insertion is illustrated by two filled squares, each
        indicating a quark bilinear. \label{fig:bs-diagrams}}
\end{figure}

In \autoref{fig:bs-diagrams}, we show examples of SMEFT four-quark operators contributing to $\mathcal{C}^{sbsb}_1$, both at tree level (left) and at one-loop level (right).
At tree level, only four-quark operators contribute to $\Delta m_s$. The matching condition for the SMEFT coefficients onto $\mathcal{C}^{sbsb}_1$ schematically reads
\begin{align}\label{eq:sbsb-tree}
    \mathcal{C}^{sbsb}_{1,\text{tree}}
        & \sim V_{k2}^\ast V_{k 3} V_{32}^\ast V_{33} \left( (C_{qq}^{(+)})_{33kk} + (C_{qq}^{(+)})_{3kk3} \right)\\\nonumber
        & = (V_{ts}^\ast V_{tb})^2\, \bbb^{(+)} y_t^4
          + \mathcal{O}(y_b)\,.
\end{align}
At the one-loop level, $\mathcal{C}^{sbsb}_1$ is sensitive to further SMEFT contributions~\cite{Dekens:2019ept},
which we include in our analysis. The four-quark operators $O_{qq}^{(1)}$ and $O_{qq}^{(3)}$, which enter
$\mathcal{C}^{sbsb}_1$ at tree level in two specific combinations, contribute at one-loop level with different
gauge structures.
In the MFV-SMEFT, these two types of structures yield contributions of the form
\begin{align}\label{eq:sbsb-loop}
(\bar s\gamma_\mu P_L b)(\bar t \gamma^\mu P_L t) :\   \mathcal{C}_{1,\text{loop}}^{sbsb} & \sim V_{k2}^\ast V_{k 3} V_{32}^\ast V_{33} \,(C_{qq}^{(-)})_{33kk}\\\nonumber
&  = (V_{ts}^\ast V_{tb})^2\left(\aat^{(-)} + \bat^{(-)} y_t^2 + \bab^{(-)} y_t^2 + \bbb^{(-)} y_t^4\right)\\\nonumber
& \equiv (V_{ts}^\ast V_{tb})^2 F_{qq}^{(-)},\\\nonumber
(\bar s\gamma_\mu P_L t)(\bar t \gamma^\mu P_L b) :\   \mathcal{C}_{1,\text{loop}}^{sbsb} & \sim V_{k2}^\ast V_{k 3} V_{32}^\ast V_{33} \,(C_{qq}^{(3)})_{3kk3}\\\nonumber
& = (V_{ts}^\ast V_{tb})^2\left(\aa^{(3)} + \ba^{(3)} y_t^2 + \bab^{(3)} y_t^2 + \bbb^{(3)} y_t^4 \right)\\\nonumber
& \equiv (V_{ts}^\ast V_{tb})^2 F_{qq}^{(3)},
\end{align}
up to contributions of $\mathcal{O}(y_b^2)$ or smaller. The flavor structures $F_{qq}^{(-)}$ and $F_{qq}^{(3)}$ appearing in \autoref{eq:sbsb-loop} are also probed
by $\mathcal{C}^{sb\mu\mu}_{10}$ in $\mathcal{B}(B_s\to \mu^+\mu^-)$~\cite{Bruggisser:2021duo},
but with different relative contributions. Combining $\Delta m_s$ and $\mathcal{B}(B_s\to \mu^+\mu^-)$ allows us to distinguish between these two structures, but not to disentangle the individual flavor parameters within $F_{qq}^{(-)}$ and $F_{qq}^{(3)}$.

Combining tree-level and one-loop contributions, four-quark operators in the MFV-SMEFT contribute to $\mathcal{C}_1^{sbsb}$ as
\begin{align}
\mathcal{C}_1^{sbsb} \sim (V_{ts}^\ast V_{tb})^2\,\Big[\bbb^{(+)} y_t^4 + \frac{g^2}{16\pi^2} \left(L_{qq}^{(-)} F_{qq}^{(-)} + L_{qq}^{(3)} F_{qq}^{(3)}\right)\Big],
\end{align}
where $L_{qq}^{(-)}$ and $L_{qq}^{(3)}$ are loop functions.

The two-quark operators $O_{\phi q}^{(1)}$ and $O_{\phi q}^{(3)}$ modify the $tsW$, $tbW$ and $bsZ$ couplings
in one-loop diagrams with internal electroweak gauge bosons and top quarks. Their contributions match onto $\mathcal{C}_1^{sbsb}$ as
\begin{align}\label{eq:sbsb-loop-2q}
tbW,\,tsW: \quad  \mathcal{C}_{1,\text{loop}}^{sbsb} & \sim 2V_{ts}^\ast V_{tb} \left(V_{k2}^\ast V_{33}\,(C_{\phi q}^{(3)})_{k3} + V_{32}^\ast V_{k3}\,(C_{\phi q}^{(3)})_{3k} \right)\\\nonumber
& = (V_{ts}^\ast V_{tb})^2\,4(C_{\phi q}^{(3)})_{33} = (V_{ts}^\ast V_{tb})^2 4\left(a^{(3)} + b^{(3)}y_t^2\right)\\\nonumber
bsZ:\quad  \mathcal{C}_{1,\text{loop}}^{sbsb} & \sim V_{ts}^\ast V_{tb} V_{k2}^\ast V_{k3}(C_{\phi q}^{(+)})_{kk} \\\nonumber
& = (V_{ts}^\ast V_{tb})^2\, b^{(+)} y_t^2.
\end{align}
The full contribution of two-quark operators to $\mathcal{C}_1^{sbsb}$ in the MFV-SMEFT reads
\begin{align}
\mathcal{C}_{1,\text{loop}}^{sbsb} \sim (V_{ts}^\ast V_{tb})^2\frac{g^2}{16\pi^2} \Big[ L_{tbW} \left(a^{(3)} + b^{(3)}y_t^2\right)  + L_{bsZ}\, b^{(+)}y_t^2 \Big],
\end{align}
where $L_{tbW}$ and $L_{bsZ}$ are loop functions. 

In contrast to the $sb$ and $sb\mu\mu$ observables, the theory prediction for $\Delta m_s$ is \emph{linear} in the WET coefficients. This difference is due to the fact that the mass difference
is directly sensitive to the \emph{amplitude} in $B_s - \overline{B}_s$ mixing, rather than the squared amplitude as for decay rates. As explained at the beginning of \autoref{sec:flavor}, we do not include SMEFT contributions to the amplitude beyond $\mathcal{O}(\Lambda^{-2})$. In particular, we do not consider double insertions of the two-quark operators $O_{\phi q}^{(w)}$ in the matrix element for $B_s-\overline{B}_s$ mixing, which would generate a tree-level matching contribution to $\mathcal{C}_1^{sbsb}$ of $\mathcal{O}(\Lambda^{-4})$.

In~\autoref{tab:sbsb:deltam_s}, we provide our numerical result for $\Delta m_s$ in terms of the MFV-SMEFT parameters at the reference scale $\mu_0 = 2.4\,$TeV. The tree-level contribution of the four-quark combination $\bbb^{(+)} = \bbb^{(1)} + \bbb^{(3)}$ exceeds the loop-induced SM contribution by about a factor of $10$. As expected, $\Delta m_s$ is indeed a very sensitive probe of double flavor breaking in four-quark couplings. The (negligibly) small difference between the contributions of $\bbb^{(1)}$ and $\bbb^{(3)}$ is due to loop contributions in the matching.

Besides the strong sensitivity to double flavor breaking, $\Delta m_s$ also receives significant contributions from two-quark operators and other four-quark coefficients. These contributions consist of two parts: matching at one-loop level, see \autoref{eq:sbsb-loop} and \autoref{eq:sbsb-loop-2q}; and RG mixing with the four-quark structure \autoref{eq:sbsb-tree} that matches onto $\mathcal{C}_1^{sbsb}$ at tree level. Contributions from operator mixing depend strongly on the reference scale $\mu_0$. Similar effects have also been observed in $B_s \to \mu^+\mu^-$~\cite{Bruggisser:2021duo}. The scale dependence is a consequence of the mismatch between one-loop matching and LL RG evolution. Performing the RG evolution at NLL would reduce this scale dependence.

\section{Weak scale: top and $Z-$pole observables}\label{sec:weak-scale}
Around the weak scale, a large number of collider observables allow us to probe the operator set from \autoref{eq:operators:4q} and \autoref{eq:operators:2q}. In SMEFT, high-energy observables have the same structure as the flavor observables from \autoref{eq:decay-rate-smeft}. They are sesquilinear polynomials in the flavor parameters from \autoref{eq:2q-flavor} and \autoref{eq:4q-flavor}.

We focus on top and $Z-$pole observables, which probe different directions in flavor space. This set of observables is sensitive to all flavor parameters in \autoref{eq:2q-flavor} and \autoref{eq:4q-flavor}. However, it can only probe the combinations $\aa + \ba y_t^2$ and $\aat + \bat y_t^2$, which are typical for top-quark couplings, but not the individual parameters $\aa,\ba$ or $\aat,\bat$.

In \autoref{sec:ttbar}, we review top observables that are sensitive to the four-quark operators $O_{qq}^{(w),33ii},O_{qq}^{(w),3ii3}$ and two-quark operators $O_{\phi q}^{(w),11},O_{\phi q}^{(w),33}$ from a previous analysis~\cite{Bruggisser:2021duo}, which builds on a global analysis of the top sector from Ref.~\cite{Brivio:2019ius}. We adopt the results of this analysis for this work. Other global fits of the top sector in SMEFT~\cite{Buckley:2015lku,Hartland:2019bjb} should lead to similar results. In \autoref{sec:ttbb}, we investigate $t\bar{t} b\bar{b}$ and $t\bar{t}t\bar{t}$ production at the LHC, which in addition probe the operators $O_{qq}^{(w),3333}$ that involve only quarks from the third generation. In \autoref{sec:z-pole}, we explore $Z-$pole observables as an alternative to probe two-quark operators and $O_{qq}^{(w),3333}$.

\subsection{Top-antitop, $t\bar{t}Z$, $t\bar{t}W$ and single-top production}\label{sec:ttbar}
\noindent
At the LHC top-quark production has been investigated in various channels and kinematic distributions. This provides us with a large number of observables, which allow us to probe many parameter directions in the SMEFT. In general, top observables can be ordered in two groups: hadronic $t\bar{t}$ production is sensitive to four-quark operators; electroweak top production is in addition sensitive to two-quark operators. Observables in these two groups probe the following flavor parameter combinations in the MFV-SMEFT:
\begin{align}
    t\bar t \text{ production}: & \quad \aa^{(w)} + \ba^{(w)},\ \aat^{(w)} + \bat^{(w)}\\\nonumber
    \text{single-top},\ t\bar{t}Z,\ t\bar{t}W: & \quad a^{(w)},\ b^{(w)},\ \aa^{(w)} + \ba^{(w)},\ \aat^{(w)} + \bat^{(w)}\,.
\end{align}
Altogether, these are $4 + 4 = 8$ directions in flavor space. In a previous analysis, we have analyzed the SMEFT contributions to top observables and the impact on global fits in detail. We refer the reader to Ref.~\cite{Bruggisser:2021duo} for more information.

\subsection{$t\bar{t}b\bar{b}$ and $t\bar{t}t\bar{t}$ production}\label{sec:ttbb}
\noindent As we see in \autoref{sec:ttbar}, $t\bar t$ production resolves the full flavor space of four-quark operators with two light and two heavy quarks, namely  $O_{qq}^{(w),33ii}$ and $O_{qq}^{(w),3ii3}$. To probe operators $O_{qq}^{(w),3333}$ with four third-generation quarks, one has to resort to $t\bar{t}t\bar{t}$ or $t\bar{t}b\bar{b}$ production. These processes are particularly interesting, because they probe double flavor breaking in the MFV-SMEFT at tree level. Due to the different weak isospin currents of the involved quarks, $t\bar{t}t\bar{t}$ or $t\bar{t}b\bar{b}$ production probe different combinations of the Wilson coefficients,
\begin{align}
t\bar{t}t\bar{t}: \ (C_{qq}^{(1)})_{3333} + (C_{qq}^{(3)})_{3333} \supset \bbb^{(+)}, \quad
t\bar{t}b\bar{b}: \ (C_{qq}^{(1)})_{3333} - (C_{qq}^{(3)})_{3333} \supset \bbb^{(-)}.
\end{align}
In \autoref{sec:sbsb}, we have shown that $B_s$--$\overline{B}_s$ mixing is very sensitive to double flavor breaking, $\bbb^{(+)}$, in the MFV-SMEFT. We have checked with simulations that existing measurements of $t\bar{t}t\bar{t}$ production at the LHC~\cite{CMS:2019rvj,ATLAS:2020hpj,ATLAS:2021kqb} are much less sensitive to $\bbb^{(+)}$. This can also be inferred from a recent detailed analysis of four-top production in SMEFT~\cite{Aoude:2022deh}.

On the contrary, $t\bar{t}b\bar{b}$ production is the main process to probe the orthogonal combination $\bbb^{(-)}$. We therefore include $t\bar{t}b\bar{b}$ production in our analysis. Effective couplings of four heavy quarks can also be probed through loop effects in top-antitop production~\cite{Degrande:2020evl}, but with less sensitivity than in $t\bar{t}b\bar{b}$ and $t\bar{t}t\bar{t}$ production.

For $t\bar t b\bar{b}$ production at the 13-TeV LHC, the total cross section in the MFV-SMEFT reads
\begin{align}
    \sigma_{t\bar t b\bar b} = \sigma_{t\bar t b\bar b}^{\rm SM} + \bigg[0.029\,\bbb^{(-)}\left(\frac{\rm TeV}{\Lambda}\right)^2 + 0.067 \left(\bbb^{(-)}\right)^2\left(\frac{\rm TeV}{\Lambda}\right)^4\bigg]\text{pb}.
\end{align}
Here and in our numerical analysis we neglect contributions from operators with two third-generation quarks, which are much more strongly constrained by top-antitop and electroweak top production, see \autoref{sec:ttbar}.

Measuring $t\bar{t}b\bar{b}$ production at the LHC is challenging. The CMS collaboration has extracted the total $t\bar{t}b\bar{b}$ cross section from Run 2 data, assuming SM backgrounds~\cite{CMS:2020grm}. Using this measurement we find a bound on $\bbb^{(-)}$ in a single-parameter fit at $\Delta \chi^2 = 3.84$,
\begin{align}\label{eq:ttbb-bound}
 -\frac{6.5}{\rm{TeV}^2} < \frac{\bbb^{(-)}}{\Lambda^2} < \frac{6.1}{\rm{TeV}^2}\,.
\end{align}

\subsection{$Z-$pole observables}\label{sec:z-pole}
\noindent LEP observables at the $Z$ resonance have not only been precisely measured, but are also very sensitive to modifications of the electroweak fermion couplings in SMEFT~\cite{Barbieri:1999tm,Pomarol:2013zra,Falkowski:2014tna}. Among the operators from \autoref{eq:operators:4q} and \autoref{eq:operators:2q}, $Z-$pole observables probe mostly $O_{\phi q}^{(1)}$ and $O_{\phi q}^{(3)}$ at tree level. One-loop corrections are interesting because they probe four-quark operators with heavy quarks through top loops ~\cite{Hartmann:2016pil,Dawson:2019clf,Dawson:2022bxd}. In particular, modifications of the $Z$ couplings to bottom quarks are sensitive to the parameter $\bbb^{(-)}$, which is only loosely constrained by $t\bar{t}b\bar{b}$ production at the LHC.

Global fits of electroweak observables in SMEFT with a general flavor structure have been performed at tree level ~\cite{Han:2005pr,Efrati:2015eaa,Falkowski:2019hvp} and for flavor-universal couplings also at one-loop level~\cite{Liu:2022vgo}. We will see that $Z-$pole observables have a significant impact on our global fit, even though they are not indispensable to fully resolve the flavor structure in the MFV-SMEFT.

At leading order, the Lagrangian for $Z$ boson couplings to quarks in the MFV-SMEFT reads~\cite{Dawson:2019clf}
\begin{align}
       \mathcal{L} & = 2 M_Z (\sqrt{2}G_F)^\frac{1}{2}Z_{\mu}\bigg[\left(g_L^u -\frac{v^2}{2\Lambda^2} (C_{\phi q}^{(-)})_{kk}\right)\left(\overline{u}_L^k \gamma^\mu u_L^k\right)\\\nonumber
    & \qquad\qquad\qquad\qquad\quad + \left(g_L^d -\frac{v^2}{2\Lambda^2} (C_{\phi q}^{(+)})_{kk}\right)\left(\overline{d}_L^k \gamma^\mu d_L^k\right)\bigg]\\\nonumber
    & = 2 M_Z(\sqrt{2}G_F)^\frac{1}{2}Z_{\mu}\bigg[\left(g_L^u -\frac{v^2}{2\Lambda^2} a^{(-)}\right)\sum_k \left(\overline{u}_L^k \gamma^\mu u_L^k\right)
    -\frac{v^2}{2\Lambda^2} b^{(-)}y_t^2\Big(\overline{t}_L \gamma^\mu t_L\Big)\\\nonumber
    & \qquad\qquad\qquad\qquad\quad + \left(g_L^d -\frac{v^2}{2\Lambda^2} a^{(+)}\right)\sum_k\left(\overline{d}_L^k \gamma^\mu d_L^k\right) -\frac{v^2}{2\Lambda^2} b^{(+)} y_t^2\Big(\overline{b}_L \gamma^\mu b_L\Big) \bigg],
\end{align}
choosing the electroweak input parameters as $\{\alpha,M_Z,G_F\}$ and neglecting CKM-suppressed contributions. The SM couplings are $g_L^u = \frac{1}{2} - \frac{2}{3}s_W^2$ and $g_L^d = -\frac{1}{2} + \frac{1}{3}s_W^2$; $v$ is the vacuum expectation value of the Higgs field.

To demonstrate the role of $Z-$pole observables in probing the flavor structure of Wilson coefficients, we select two observables that are particularly sensitive to flavor breaking and extremely well measured: the total width of the $Z$ boson, $\Gamma_Z$, and the normalized partial width for $Z\to b\bar{b}$ decays, $R_b = \Gamma_{b\bar{b}}/\Gamma_Z$. At tree level, the leading contributions to the partial and total decay width are (neglecting quark masses)
\begin{align}
    \Gamma_{b\bar{b}}^{\rm tree} & = \Gamma_{b\bar{b}}^{\rm SM} + \big(a^{(+)} + b^{(+)}y_t^2\big)\frac{v^2}{\Lambda^2}\,\delta \Gamma_Z^d \\\nonumber
    \Gamma_Z^{\rm tree} & = \Gamma_Z^{\rm SM} + 2 a^{(-)} \frac{v^2}{\Lambda^2}\,\delta \Gamma_Z^u + \big(3a^{(+)} + b^{(+)}y_t^2\big)\frac{v^2}{\Lambda^2}\,\delta \Gamma_Z^d\,.
\end{align}
Here $\Gamma_{b\bar{b}}^{\rm SM}$ and $\Gamma_Z^{\rm SM}$ are the decay widths in the Standard Model, and $\delta \Gamma_Z^u$ and $\delta \Gamma_Z^d$ are the (normalized) contributions from SMEFT operator interference with the SM amplitudes for $Z\to u^i\bar{u}^i$ and $Z\to d^k\bar{d}^k$, respectively. Quadratic contributions in the SMEFT coefficients of $\mathcal{O}\big(a^2\big)$ etc. have been neglected, but are included in our numerical analysis. Among the three contributing parameters $\{a^{(1)},a^{(3)},b^{(+)}\}$, $a^{(3)}$ is constrained at the permille level by precision tests of CKM unitarity in weak charged currents~\cite{FlaviaNetWorkingGrouponKaonDecays:2010lot}.

\begin{figure}[t]
    \centering
    \includegraphics[width=\textwidth]{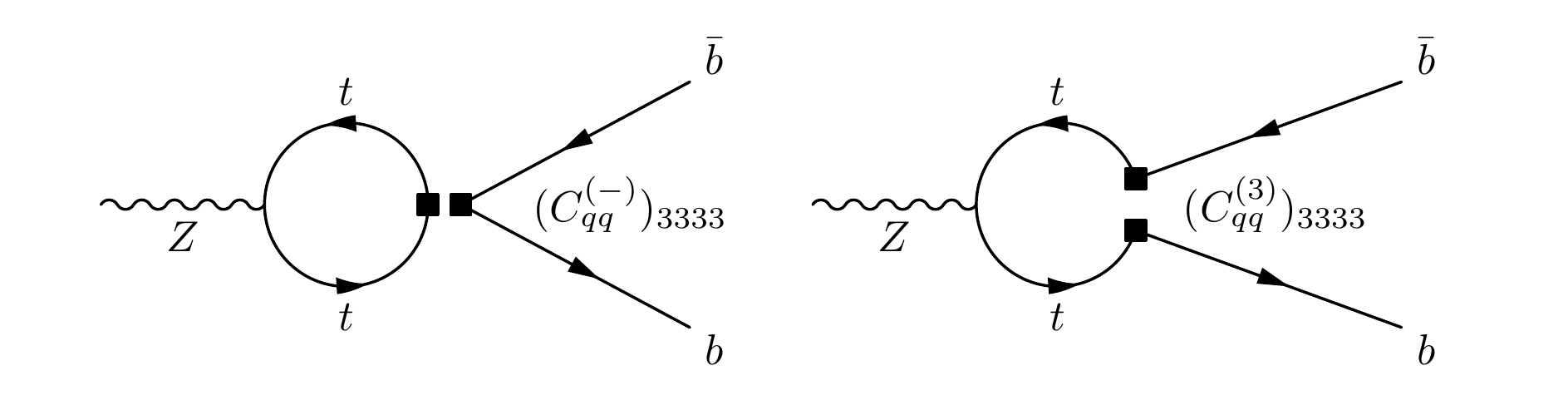}
    \caption{Illustration of two different insertions of four-quark SMEFT operators in $Z\to b\bar{b}$. Left: $s$-channel topology, with insertion of $(C_{qq}^{(-)})_{3333}$.
    Right: $t$-channel topology, involving a charged current with insertion of $(C_{qq}^{(3)})_{3333}$.
    The operator insertion is illustrated by two filled squares, each indicating a quark bilinear. \label{fig:Zpole}}
\end{figure}

Four-quark operators affect $Z$ decays through loop contributions. Here we are particularly interested in couplings of four heavy quarks, $(C_{qq}^{(w)})_{3333}$, which contribute to $Z\to b\bar{b}$ decays via top-quark loops. In \autoref{fig:Zpole}, we show two possible types of operator insertions.
In the MFV-SMEFT, these two different loop structures contribute as
\begin{align}\label{eq:zbb-loop}
    (\bar b\gamma_\mu P_L b)(\bar t \gamma^\mu P_L t) :\ \Gamma_{b\bar{b}}^{\rm loop} & \sim |V_{k3}|^2 \,(C_{qq}^{(-)})_{33kk}\\\nonumber
& = \aa^{(-)} + \ba^{(-)} + |V_{tb}|^2 F_{qq}^{(-)}\\\nonumber
& \approx \aab^{(-)} + 2\bab^{(-)} + \bbb^{(-)} = (C_{qq}^{(-)})_{3333}\\\nonumber
(\bar b\gamma_\mu P_L t)(\bar t \gamma^\mu P_L b) :\   \Gamma_{b\bar{b}}^{\rm loop} & \sim |V_{k3}|^2 \,(C_{qq}^{(3)})_{3kk3}\\\nonumber
& = \aat^{(3)} + \bat^{(3)} + |V_{tb}|^2 F_{qq}^{(3)}\\\nonumber
& \approx \aab^{(3)} + 2\bab^{(3)} + \bbb^{(3)} = (C_{qq}^{(3)})_{3333}.
\end{align}
The parameter combinations $F_{qq}^{(-)}$ and $F_{qq}^{(3)}$ are the same as for the loop contributions to meson mixing, see \autoref{eq:sbsb-loop}, and to $B_s\to \mu^+\mu^-$~\cite{Bruggisser:2021duo}. Compared to down-quark flavor-changing neutral currents, flavor-conserving currents like $Z\to b\bar{b}$ receive additional four-quark contributions, see \autoref{eq:zbb-loop}. As we will see in \autoref{sec:global}, global fits are sensitive to loop-induced four-quark contributions, even though they are numerically much smaller than tree-level effects from two-quark couplings.

For the numerical analysis of the $Z-$pole observables, we use the one-loop predictions for general flavor structures from Ref.~\cite{Dawson:2022bxd} and the theory uncertainties from Ref.~\cite{Dawson:2019clf}. Using our MFV parametrization, we compare these predictions to precision measurements at LEP~\cite{ALEPH:2005ab},
\begin{align}
    \Gamma_Z & = 2.4952 \pm 0.0023\,\text{GeV}\\\nonumber
    R_b & = 0.21629 \pm 0.00066.
\end{align}
The resulting bounds on the flavor parameters will be discussed in \autoref{sec:global}.

\section{TeV scale: dijets}\label{sec:dijets}
As mentioned in \autoref{sec:weak-scale}, top-quark observables only probe the sum of flavor-universal and flavor-breaking structures of effective four-quark interactions, parametrized by $\aa + \ba y_t^2$ and $\aat + \bat y_t^2$. Dijet production at the LHC is an excellent probe of light-quark couplings, that is, of $\aa$ and $\aat$ individually. In combination, top and dijet observables can pin down the possible amount of flavor breaking, $\ba$ and $\bat$, in four-quark interactions.

Angular correlations of two hard jets produced via $pp\to jj + X$ are among the most precise QCD tests. It is well known that they are also very sensitive to four-quark interactions in extensions of the Standard Model~\cite{Bai:2011ed,Haisch:2011up,Alte:2017pme}. At the Tevatron and the LHC, dijet angular correlations have been measured in terms of rapidity differences $|y_1 - y_2|$, where $y_1$ and $y_2$ are the rapidities of the two jets with the highest transverse momenta in each event. The results are reported in bins of the dijet invariant mass, $M_{jj}$, normalized to the cross section in the respective bin, $\sigma(M_{jj})$:
\begin{align}\label{eq:dijet-dist}
\frac{1}{\sigma(M_{jj})}\frac{d\sigma(M_{jj})}{d\chi},\quad \chi = \exp\big(|y_1 - y_2|\big) \in [1,\infty].
\end{align}
In QCD, dijet production is dominated by Rutherford scattering, which becomes singular in the forward region. The variable $\chi$ is chosen to subtract the Rutherford singularity, such that the QCD prediction for \autoref{eq:dijet-dist} is mostly flat, $(d\sigma/d\chi)_{\rm QCD} \propto$ const.

In SMEFT, the four-quark operators $O_{qq}^{(1)}$ and $O_{qq}^{(3)}$ modify the dijet distributions, so that\footnote{The two-quark operators from \autoref{eq:operators:2q} can only contribute to dijet production via $Z$ or $W$ exchange. We neglect such electroweak contributions.}
\begin{align}\label{eq:dijet-dist-smeft}
\frac{d\sigma}{d\chi} = \left(\frac{d\sigma}{d\chi}\right)_{\rm QCD} + \sum_{w=\{1,3\}}\!\!\frac{C_{qq}^{(w)}}{\Lambda^2} \left(\frac{d\sigma_{w}}{d\chi}\right)_{\rm int} + \sum_{v,w=\{1,3\}}\!\!\!\!\frac{C_{qq}^{(v)}C_{qq}^{(w)}}{\Lambda^4}\left(\frac{d\sigma_{vw}}{d\chi}\right)_{\rm SMEFT}.
\end{align}
As in \autoref{eq:decay-rate-smeft}, the labels `int' and `SMEFT' stand for contributions from SM-SMEFT and SMEFT-SMEFT amplitude interference. For the dominant partonic amplitude $qq\to qq$, the SM-SMEFT interference scales as~\cite{Haisch:2011up}
\begin{align}
\left(\frac{d\sigma_w}{d\chi}\right)_{\rm int}\!\!(qq\to qq) \propto -\frac{1}{\chi}\frac{C_{qq}^{(w)}}{\Lambda^2}.
\end{align}
Normalized to the total rate, this results in an enhancement for $\chi \to 1$ relative to the QCD prediction -- or a depletion, depending on the sign of $C_{qq}^{(w)}$. For the SMEFT-SMEFT interference, the effect is even more pronounced.

In terms of the dijet invariant mass, the distributions scale as
\begin{align}\label{eq:dijet-mjj}
    \left(\frac{d\sigma(M_{jj})}{d\chi}\right)_{\rm QCD} \propto \frac{1}{M_{jj}^2},\quad \left(\frac{d\sigma(M_{jj})}{d\chi}\right)_{\rm int} \propto \frac{1}{\Lambda^2},\quad \left(\frac{d\sigma(M_{jj})}{d\chi}\right)_{\rm SMEFT} \propto \frac{M_{jj}^2}{\Lambda^4}.
\end{align}
As for many high-energy observables, the `int' and `SMEFT' contributions are UV-sensitive and increase as $M_{jj}^2/\Lambda^2$ and $M_{jj}^4/\Lambda^4$ relative to the QCD prediction.

\begin{figure}[t]
    \centering
    \includegraphics[width=\textwidth]{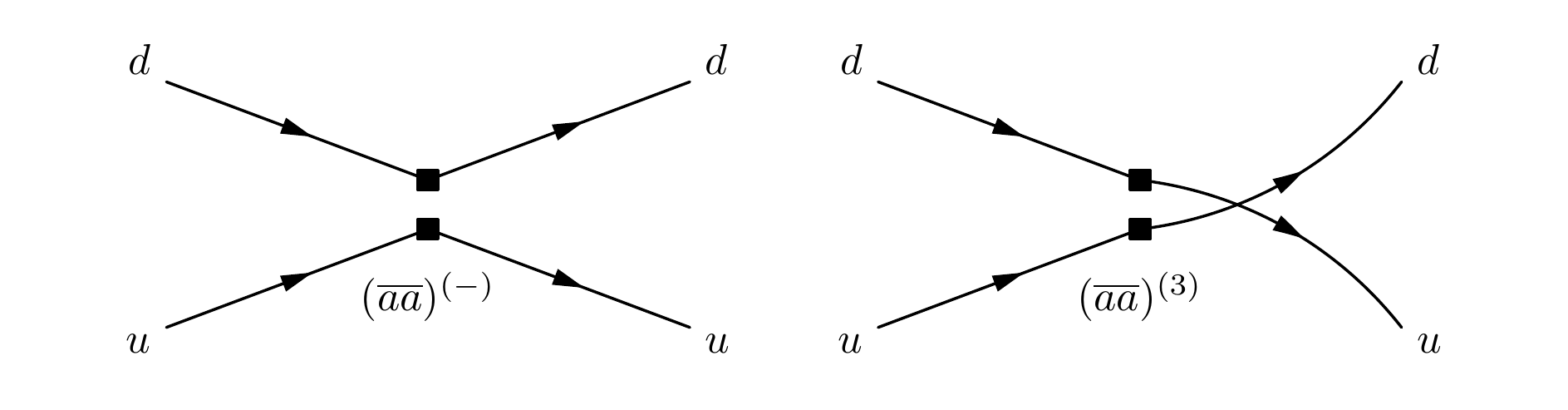}
    \caption{%
        Illustration of the two different topologies of SMEFT contributions to dijet production
        at the example of $ud \to ud$. Left: $t$-channel topology with insertion of $\aab^{(-)}$.
        Right: $u$-channel topology, involving a charged current with insertion of $\aab^{(3)}$.
        The insertion of SMEFT operators is illustrated by two filled squares, each
        indicating a quark bilinear.\label{fig:dijet}}
\end{figure}

To understand the flavor and gauge structure of the SMEFT contributions, it is instructive to look at partonic amplitudes with specific external quarks. Amplitudes with two quarks (rather than antiquarks) in the initial state, $\mathcal{A}(qq^{(')}\to qq^{(')})$, dominate due to the parton distributions inside the protons. Contributions of $O_{qq}^{(1)}$ and $O_{qq}^{(3)}$ enter with the following combinations of flavor parameters:
\begin{align}\label{eq:dijet-qq}
    \mathcal{A}(uu\to uu) & \propto (\mathcal{A}_t + \mathcal{A}_u)\,\aab^{(+)} \propto \mathcal{A}(dd\to dd)\\\nonumber
    \mathcal{A}(ud \to ud) & \propto \mathcal{A}_t\,\aab^{(-)} + \mathcal{A}_u\,\aab^{(3)}\\\nonumber
    \mathcal{A}(ud \to du) & \propto \mathcal{A}_u\,\aab^{(-)} + \mathcal{A}_t\,\aab^{(3)}.
\end{align}
Here $\mathcal{A}_t$ and $\mathcal{A}_u$ are (normalized) amplitudes with operator insertions in the $t$-channel and $u$-channel topologies. An example of these insertions is shown in \autoref{fig:dijet}.
The quark-quark amplitudes allow us to distinguish between the weak isospin structures $\aab^{(1)}$ and $\aab^{(3)}$, but leave blind directions along $\aa^{(w)} = - \aat^{(w)}$. These directions are resolved by adding SMEFT amplitudes with quark-antiquark initial states, $\mathcal{A}(q\bar{q} \to q'\bar{q}')$, despite the lower parton luminosity compared to quark-quark initial states. For instance, the amplitudes
\begin{align}\label{eq:dijet-qqb}
    \mathcal{A}(u\bar{u}\to c\bar{c}) \propto \aa^{(+)},\qquad
    \mathcal{A}(u\bar{u}\to s\bar{s}) \propto \aa^{(-)}
\end{align}
correspond to the $s$-channel topology and are only sensitive to $\aa^{(w)}$. The flavor structure $\aat^{(w)}$ would enter via $t$-channel contributions, but those are CKM and/or Yukawa-suppressed in the MFV-SMEFT. Taken together, the amplitudes in \autoref{eq:dijet-qq} and \autoref{eq:dijet-qqb} probe all 4 directions $\{\aa^{(1)},\aat^{(1)},\aa^{(3)},\aat^{(3)}\}$ in the flavor space of light-quark couplings.

For our numerical analysis of dijet angular correlations, we use the most recent measurement by the CMS collaboration, based on $35.9\,$fb$^{-1}$ of 13-TeV LHC data~\cite{CMS:2018ucw}. We include only the lowest bin in the dijet invariant mass, $2.4\,\text{TeV} < M_{jj} < 3\,\text{TeV}$, which contains the largest amount of data. As mentioned at the end of in \autoref{sec:smeft}, this motivates our choice for the reference scale $\mu_0 = 2.4\,$TeV at which we report the bounds on the flavor parameters.

\begin{figure}[t!]
    \centering
        \includegraphics[page=1,width=0.48\textwidth]{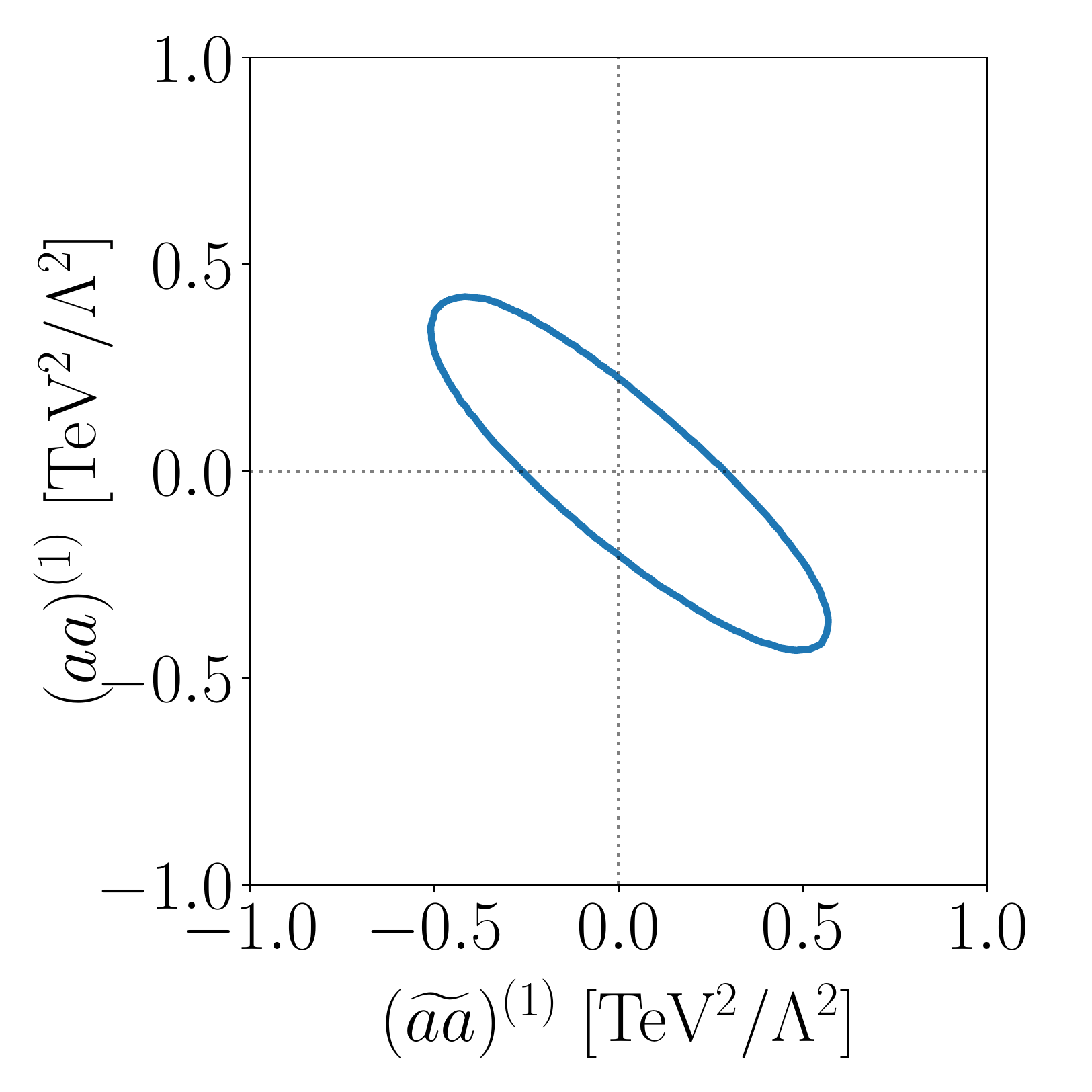}
        \quad
        \includegraphics[page=2,width=0.48\textwidth]{Figures/DijetOnlyScaled.pdf}
    \caption{Bounds on the flavor structure of the four-quark interactions $O_{qq}^{(1)}$ and $O_{qq}^{(3)}$ from a 4-parameter fit to dijet angular distributions measured at the LHC~\cite{CMS:2018ucw}. We show contours for $\Delta\chi^2=5.99$. The flavor parameters are defined at $\mu_0 = 2.4\,$TeV.}
    \label{fig:dijets}
\end{figure}

The angular distributions from \autoref{eq:dijet-dist} are reported in terms of 12 $\chi$ bins. As some of the correlations between these bins are unknown to us, we select only every second bin in the distribution, starting with the second bin, and treat them as uncorrelated. This choice is motivated by the large uncertainties affecting the first bin around $\chi = 1$.

For the QCD prediction of the distribution $d\sigma/d\chi$, we use the results of NLO simulations from Ref.~\cite{CMS:2018ucw}. To obtain the `int' and `SMEFT' contributions, see \autoref{eq:dijet-dist-smeft}, we perform simulations in SMEFT using {\tt Madgraph5\_aMC@NLO}~\cite{Alwall:2014hca} at LO QCD at parton level. A more detailed analysis should include SMEFT simulations at particle level, where the measurements are reported. For our proof-of-principle analysis, we confine ourselves to simulations at parton level.

The results of our 4-parameter fit of dijet angular distributions are shown in \autoref{fig:dijets}. The bounds are dominated by contributions from SMEFT-SMEFT interference, which are enhanced at high energies, see \autoref{eq:dijet-mjj}. As suggested by the flavor structure of the SMEFT amplitudes from \autoref{eq:dijet-qq}, partonic quark-quark contributions are very sensitive to the combination $\aab^{(w)} = \aa^{(w)} + \aat^{(w)}$ (left panel). The orthogonal direction $\aa^{(w)} - \aat^{(w)}$ is probed less, but still bounded due to quark-antiquark contributions, see \autoref{eq:dijet-qqb}. Weak triplets are more constrained than singlets (right panel). This is due to the different gauge structure, which leads to several effects in the amplitudes and their interference. Compared with most other LHC observables, dijet angular distributions are superior in their sensitivity to SMEFT couplings of four light quarks. They will play a crucial role in the global analysis of flavor structures, which we present now in \autoref{sec:global}.
 
\section{Global analysis of flavor structure}\label{sec:global}
We carry out a global fit of the observables discussed in the previous sections. For convenience, we summarise the observables' dependence on the flavor parameters at the leading order in the strong and electroweak gauge couplings:
\begin{align}\label{eq:all-observables}
   \mathcal{B}(B \to X_s \gamma)\times 10^4 & = 3.26 + 0.37\,a^{(3)} - 0.72\,b^{(3)}\\\nonumber
    \mathcal{B}(B_s \to \mu^+\mu^-)\times 10^9 & = 3.57 - 39.7\,b^{(+)} + 110.4\,\big(b^{(+)}\big)^2\\\nonumber
    \Gamma_Z & = 2.4945 + 0.0615\,a^{(1)} + 0.2236\,a^{(3)} + 0.0456\,b^{(1)} + 0.0499\, b^{(3)}\\\nonumber
    R_b & = 0.21586 + 0.01883\,a^{(1)}+0.02053\,b^{(1)}+0.02420\,b^{(3)}\\\nonumber
    \text{dijets}: & \quad\ \aa^{(w)},\ \aat^{(w)}\\\nonumber
         t\bar t \text{ production}: & \quad\ \aa^{(w)} + \ba^{(w)},\ \aat^{(w)} + \bat^{(w)}\\\nonumber
    \text{single-top},\ t\bar{t}Z,\ t\bar{t}W: & \quad\ a^{(w)},\ b^{(w)},\ \aa^{(w)} + \ba^{(w)},\ \aat^{(w)} + \bat^{(w)}\\\nonumber
    \Delta m_s\,[\text{ps}^{-1}] & = 17.28 + 156.9\,\bbb^{(+)}\\\nonumber
    \sigma_{t\bar t b\bar b}\,[\text{pb}] & = 2.4 + 0.029\,\bbb^{(-)} + 0.067\,\big(\bbb^{(-)}\big)^2.
\end{align}
The flavor parameters are defined at the scale $\mu_0 = 2.4\,$TeV and are given in units of TeV$^2/\Lambda^2$. We set $y_t = 1$ for a clearer presentation, but make no such approximation in the numerical evaluation. For the sake of clarity, for some of the observables we only show the numerically dominant SMEFT contributions. Sub-dominant linear and quadratic contributions are included in our numerical analysis. As we will discuss, they are relevant
in a global fit, since they have the power to disentangle degeneracies in the parameter space.

Our analysis extends the study conducted
in Ref.~\cite{Bruggisser:2021duo} by the following observables:
\begin{itemize}
    \item dijet angular distributions at the LHC~\cite{CMS:2018ucw};
    \item meson mixing in the $B_s$ system, $\Delta m_s$~\cite{ParticleDataGroup:2020ssz};
    \item $Z-$pole observables $\Gamma_Z$ and $R_b$~\cite{ALEPH:2005ab}.
\end{itemize}
The large data set enables us to constrain a larger set of parameters than in our previous analysis, namely the full set of two-quark and four-quark parameters. We are now able to distinguish between flavor-universal, singly flavor-breaking and doubly flavor-breaking contributions to four-quark operators, parameterized by $\aa$, $\ba$ and $\bb$. In particular, we are sensitive to the doubly flavor-breaking parameters $\bbb^{(w)}$, which were not accessible in Ref.~\cite{Bruggisser:2021duo}. We also gain resolution in probing the flavor structure of two-quark operators, parameterized by $a$ and $b$.

On the technical side, our analysis setup is the same as in Ref.~\cite{Bruggisser:2021duo}.
The only difference is the reference scale for the SMEFT coefficients, which was set to $\mu_0 = 1\,$TeV in Ref.~\cite{Bruggisser:2021duo}, but which we set to $\mu_0 = 2.4\,$TeV here to include the dijet angular distributions. To compare concrete UV models with our fit results, one should match these models onto the SMEFT amplitudes at this high scale. When presenting our numerical bounds on the flavor parameters, we set $\Lambda = 1\,$TeV.
 In this way, the results of our fit are easier to compare with other SMEFT analyses that report bounds on the Wilson coefficients at $\Lambda = \mu_0 = 1\,$TeV. For a precise comparison, one would need to evolve the relevant set of couplings to the same reference scale, including operator mixing.

We perform three separate maximum-likelihood fits of the flavor parameters from \autoref{eq:2q-flavor} and \autoref{eq:4q-flavor}, using the {\tt sfitter} software~\cite{Lafaye:2004cn,Lafaye:2007vs,Lafaye:2009vr}. We find it convenient to conduct the fit in a different basis, involving linear combinations of the
flavor parameters from \autoref{eq:2q-flavor} and \autoref{eq:4q-flavor}:
\begin{align}\label{eq:fit-basis}
   & a^{(-)} = a^{(1)} - a^{(3)},\quad a^{(3)},\quad b^{(\pm)} = b^{(1)} \pm b^{(3)}\\\nonumber
   & \aa^{(w)} \pm \ba^{(w)},\quad \aat^{(w)} \pm \bat^{(w)},\quad \bbb^{(\pm)} = \bbb^{(1)} \pm \bbb^{(3)}.
\end{align}
These linear combinations are more effectively constrained by our set of observables
 and render the evaluation of the likelihood computationally more efficient.

\begin{figure}[t]
    \centering
        \includegraphics[page=14,width=0.48\textwidth]{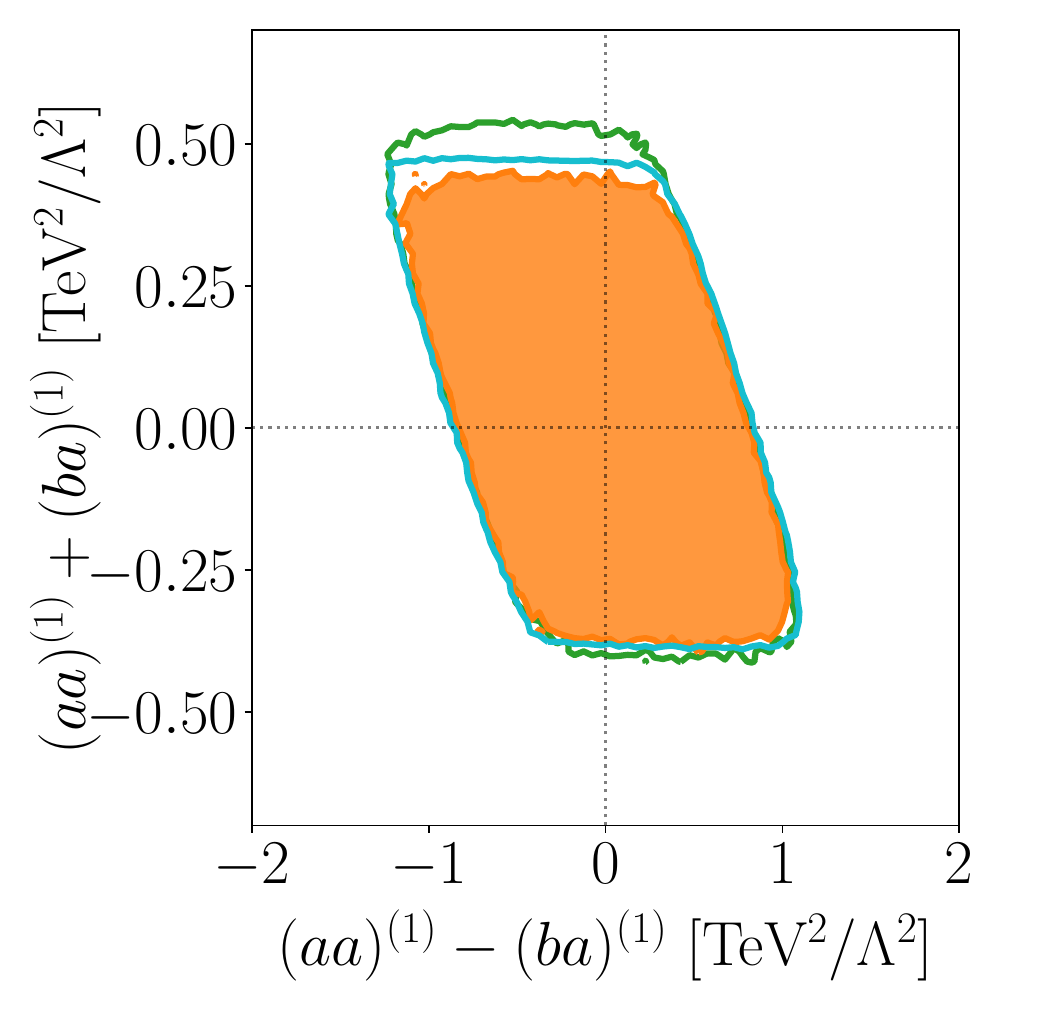}
        \quad \includegraphics[page=56,width=0.48\textwidth]{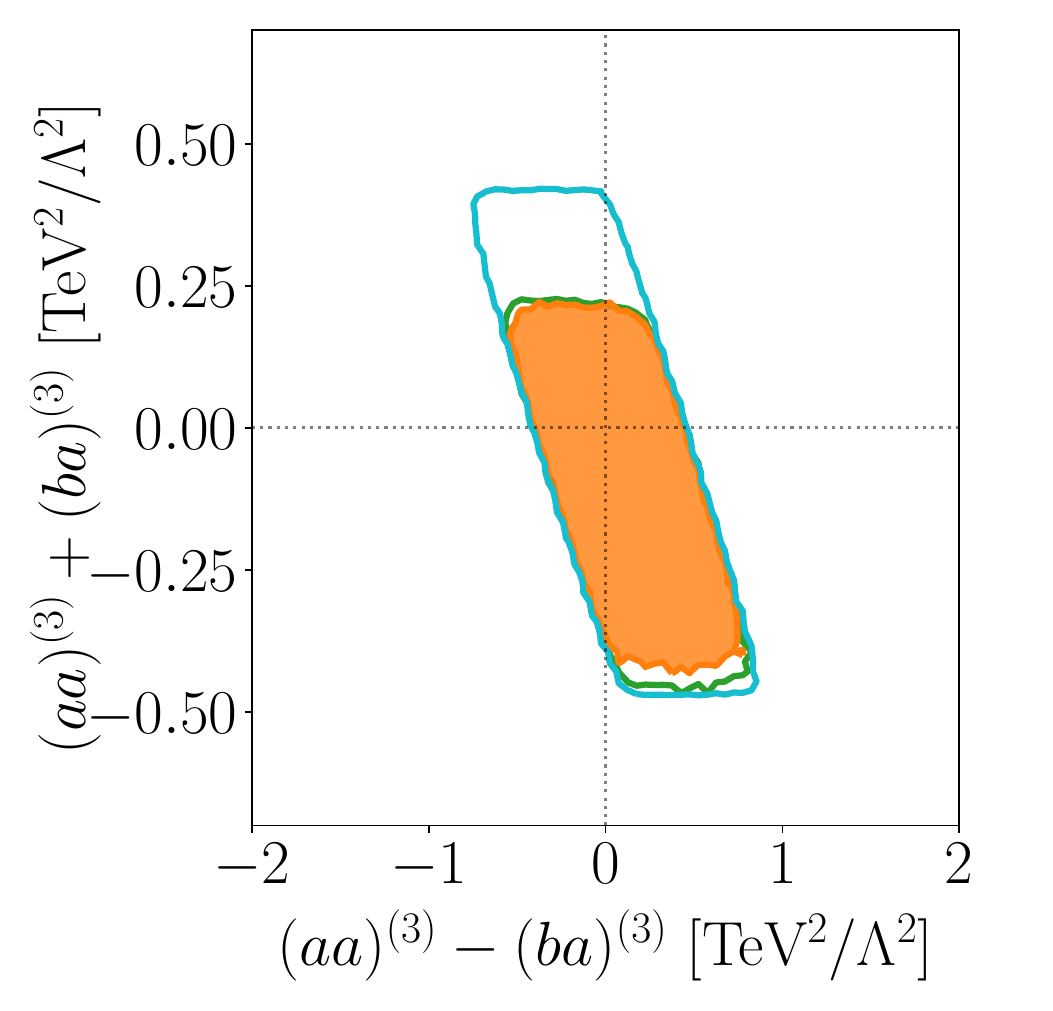}
    \caption{Bounds on flavor-universal versus singly flavor-breaking four-quark couplings, $\aa^{(w)}$ and $\ba^{(w)}$, for weak singlets (left) and weak triplets (right). Shown are likelihood contours of $\Delta \chi^2 = 5.99$ for data set 1 (cyan), set 2 (green) and set 3 (orange). The area inside the contours is in agreement with the data. The flavor parameters are defined at $\mu_0 = 2.4\,$TeV.
\label{fig:dijet-global}}
\end{figure}

\begin{figure}[t]
    \centering
        \includegraphics[page=82,width=0.48\textwidth]{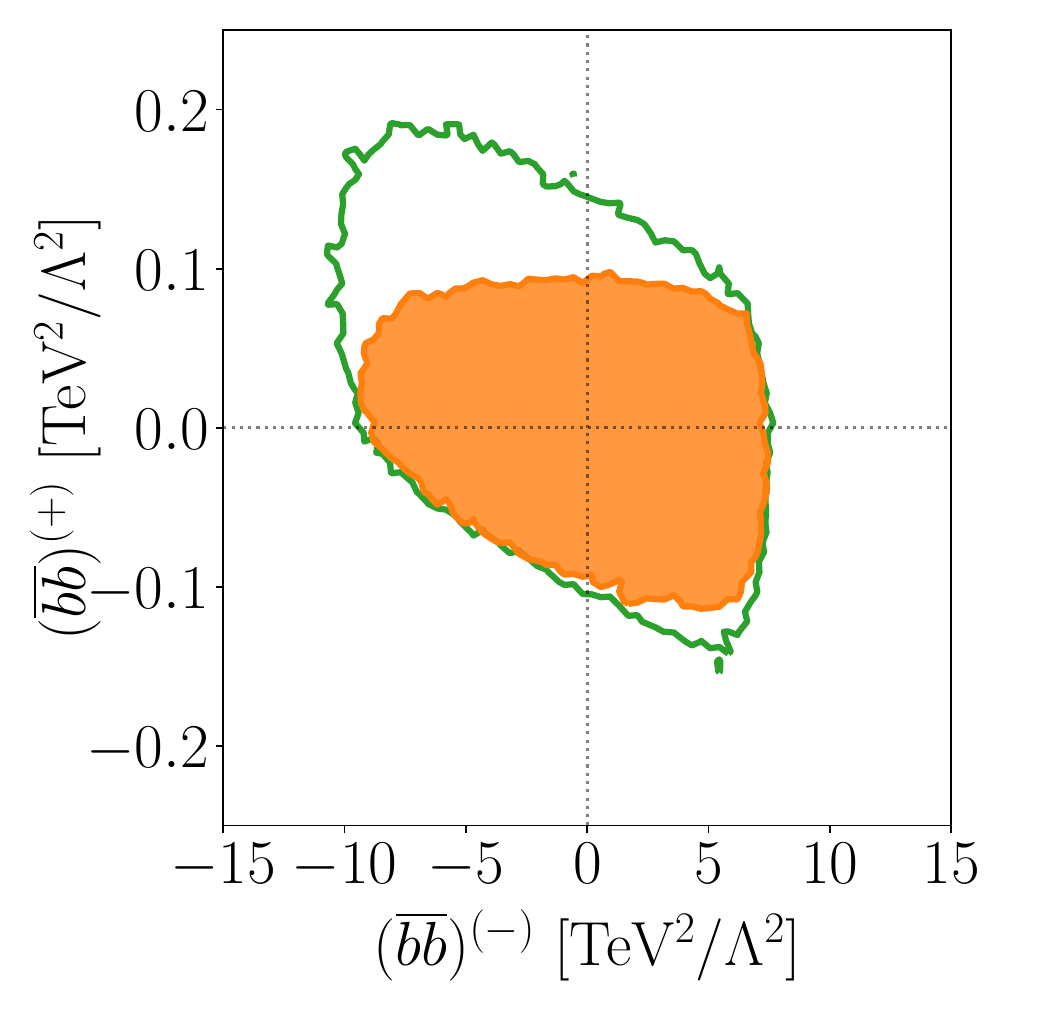}
        \quad \includegraphics[page=83,width=0.48\textwidth]{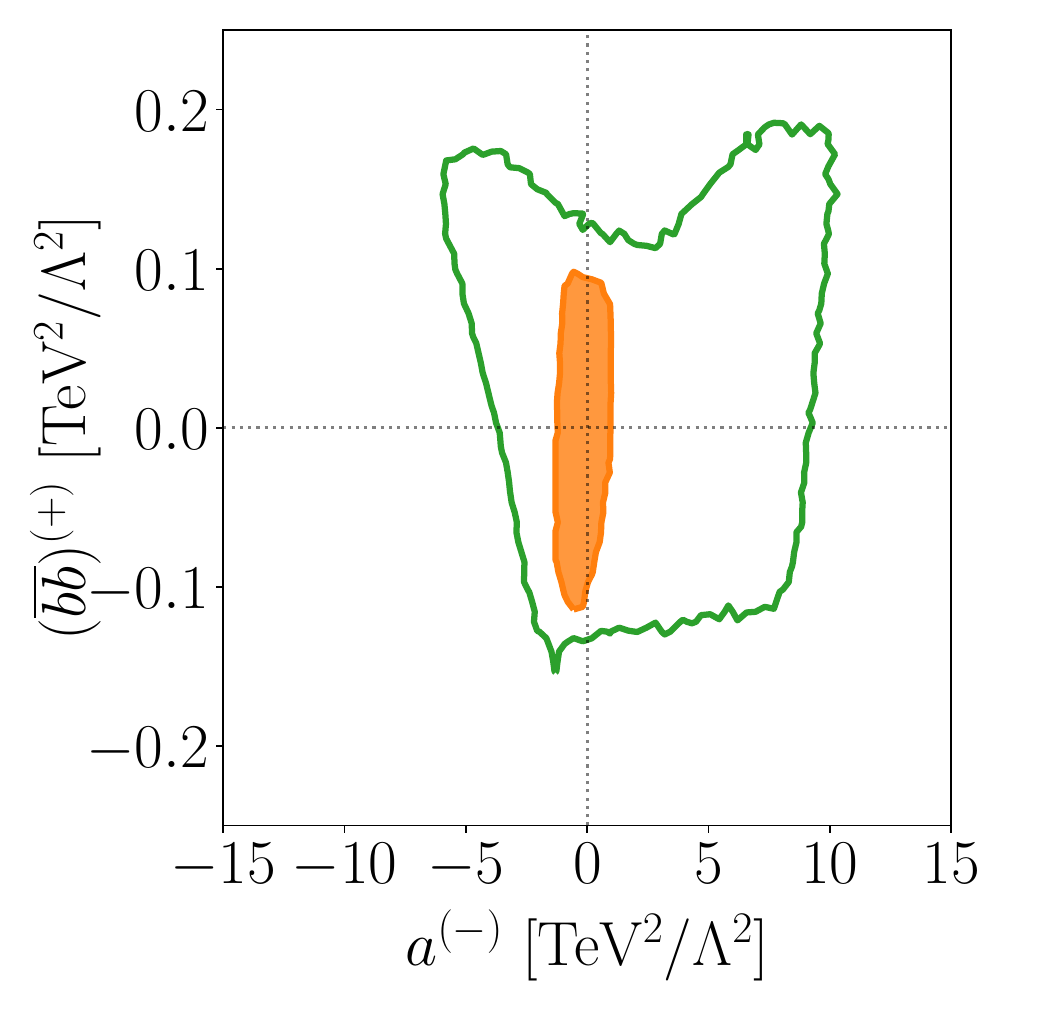}
    \caption{Bounds on double flavor breaking in four-quark couplings, $\bbb^{(w)}$. Shown are likelihood contours of $\Delta \chi^2 = 5.99$ for data set 2 (top+flavor+dijets, green) and data set 3 (including $Z-$pole data, orange). Left: weak gauge structure of $\bbb$. Right: impact of the $Z-$pole observables $\Gamma_Z$ and $R_b$ through parameter correlations. The flavor parameters are defined at $\mu_0 = 2.4\,$TeV.\label{fig:bb-bound}}
\end{figure}

\begin{figure}[t]
    \centering
        \includegraphics[page=91,width=0.48\textwidth]{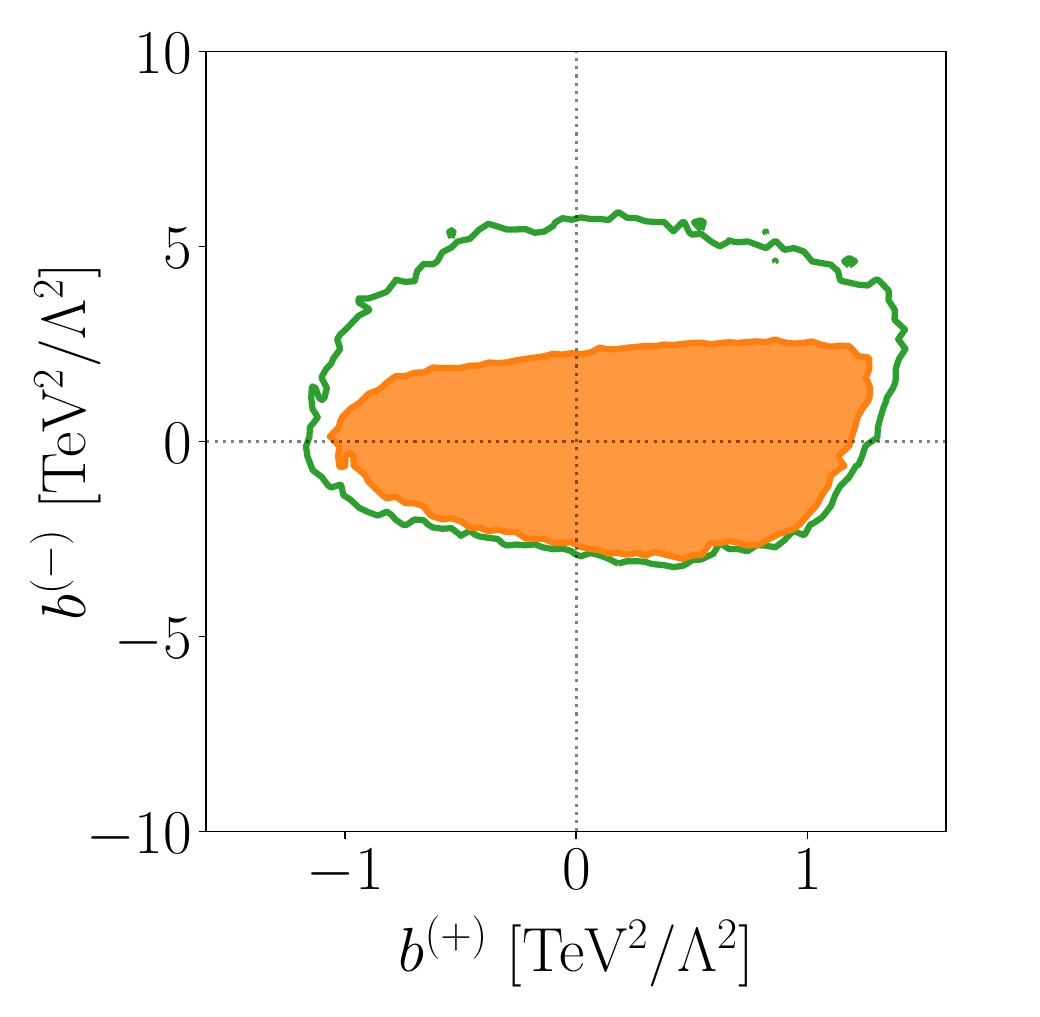}
        \quad \includegraphics[page=89,width=0.48\textwidth]{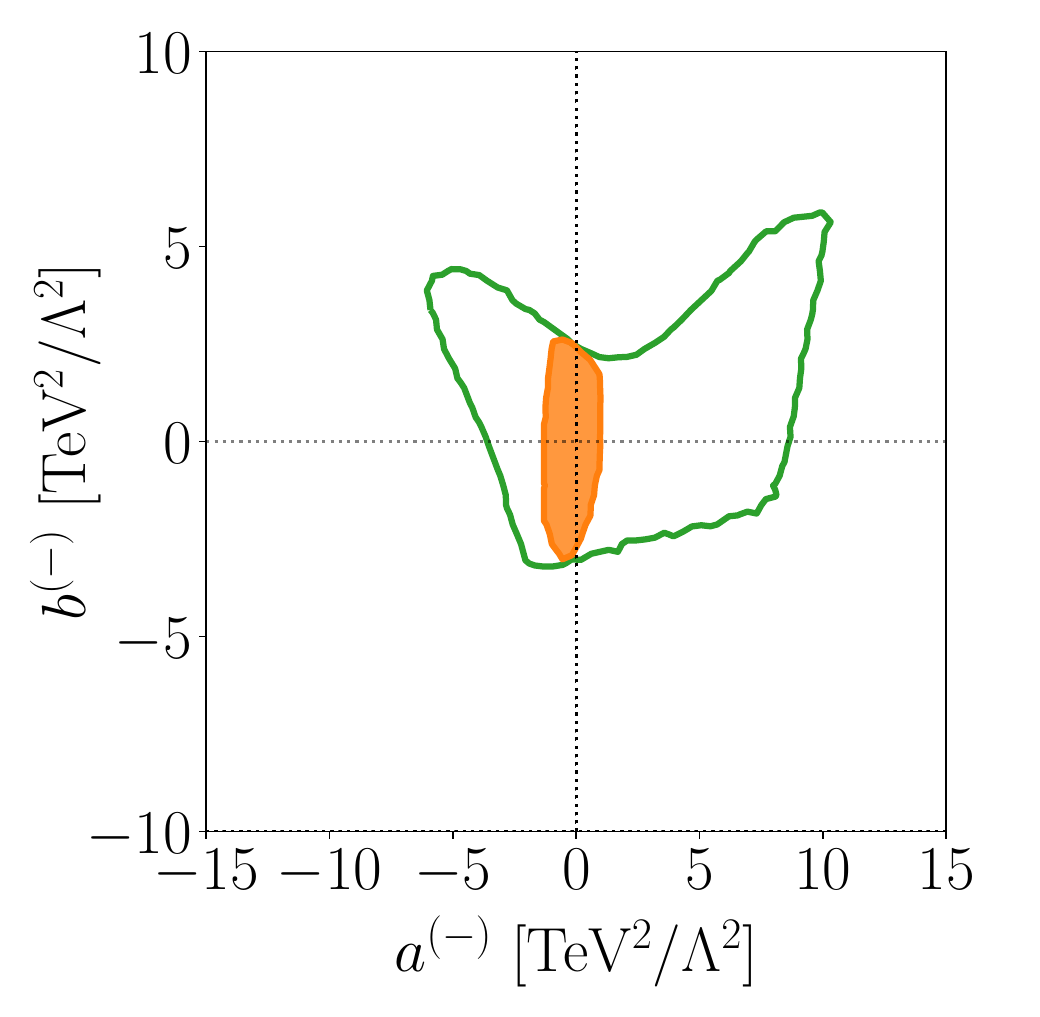}
    \caption{Bounds on flavor breaking in two-quark couplings, parametrized by $b^{(+)}$, $b^{(-)}$. Shown are likelihood contours of $\Delta \chi^2 = 5.99$ for data set 2 (top+flavor+dijets, green) and data set 3 (including $Z-$pole data, orange). Left: weak gauge structure of flavor-breaking contributions. Right: flavor-universal versus flavor-breaking couplings. The flavor parameters are defined at $\mu_0 = 2.4\,$TeV.\label{fig:correlations}}
\end{figure}

To demonstrate the impact of individual observables in the global analysis, we have performed separate fits for three data sets:
\begin{description}
    \item[data set 1] includes all observables in top-antitop production from Ref.~\cite{Brivio:2019ius} and the dijet observables discussed in \autoref{sec:dijets}. For this data set, we only fit the 8 four-quark parameters $\{\aa^{(w)},\aat^{(w)},\ba^{(w)},\bat^{(w)}\}$. The results we obtain from this fit are shown in cyan.
    \item[data set 2] includes all top observables from Ref.~\cite{Brivio:2019ius}; the dijet observables discussed in \autoref{sec:dijets}; the cross section for $t\bar{t}b\bar{b}$ production, see \autoref{sec:ttbar}; and the flavor observables $\mathcal{B}(B \to X_s \gamma)$, $\mathcal{B}(B_s \to \mu^+\mu^-)$, and $\Delta m_s$ from \autoref{sec:flavor}. We fit the full set of 14 flavor parameters in the MFV-SMEFT. The corresponding results are colored green.
    \item[data set 3] corresponds to data set 2, plus the $Z-$pole observables $\Gamma_Z$ and $R_b$. We fit the full set of 14 flavor parameters in the MFV-SMEFT. The results are colored orange.
\end{description}
The results of our analysis are shown in \autoref{tab:individual-bounds} and \autoref{fig:dijet-global}, \autoref{fig:bb-bound}, and \autoref{fig:correlations}. To obtain bounds on individual flavor parameters, we profile the full likelihood function with respect to all other parameters
and report intervals at $\Delta\chi^2 = 3.84$.
For the two-parameter plots, we profile the full likelihood with respect to all remaining parameters
and show contours at $\Delta\chi^2 = 5.99$. In the case of a fully Gaussian likelihood, these bounds would correspond to $95\%$ confidence regions. However, since our global likelihood contains non-Gaussian terms, the bounds can only be interpreted as approximations of the $95\%$ confidence level.

In \autoref{tab:individual-bounds}, we summarize the bounds on individual flavor parameters obtained from separate fits to the three data sets. Notice that data set 1 contains less parameters than data set 2 and 3; the bounds should therefore not be directly compared. The bounds on the various four-quark parameters for flavor-universal and singly flavor-breaking couplings are comparable in magnitude. The parameter directions $\aa^{(w)}+\ba^{(w)}$ are somewhat more strongly constrained than $\aa^{(w)} - \ba^{(w)}$, because the $``+"$ combination is directly aligned with the top observables, while the $``-"$ combination is only constrained in a joint fit with dijets (data set 1). Data sets 2 and 3 do not change this behavior, since top and dijet observables dominate the bounds on these parameters in the global fit.
\begin{table}[t]
\centering
\begin{tabular}{c|ccc}
\toprule
   fit   & {\bf data set 1}     & {\bf data set 2}    & {\bf data set 3} \\
parameter             & (top, dijets)        & (+ flavor)  & (+ $Z-$pole) \\
\midrule
$a^{(-)}$               &                    & {[}-5.69, 9.75{]}  & {[}-1.04, 0.80{]} \\
$a^{(3)}$               &                    & {[}-1.77, 1.27{]}  & {[}-0.08, 0.10{]} \\
$b^{(-)}$               &                    & {[}-2.63, 5.61{]}  & {[}-2.50, 2.28{]} \\
$b^{(+)}$               &                    & {[}-1.15, 1.38{]}  & {[}-1.00, 1.24{]} \\
\midrule
$\aa^{(1)}+\ba^{(1)}$   & \ {[}-0.33, 0.42{]} & {[}-0.37, 0.50{]}  & {[}-0.34, 0.43{]} \\
$\aa^{(1)}-\ba^{(1)}$   & \ {[}-1.17, 1.03{]} & {[}-1.19, 1.03{]}  & {[}-1.15, 0.99{]} \\
$\aat^{(1)}+\bat^{(1)}$ & \ {[}-0.58, 0.64{]} & {[}-0.59, 0.48{]}  & {[}-0.59, 0.44{]} \\
$\aat^{(1)}-\bat^{(1)}$ & \ {[}-1.47, 1.53{]} & {[}-1.29, 1.54{]}  & {[}-1.26, 1.49{]} \\
\midrule
$\aa^{(3)}+\ba^{(3)}$   & \ {[}-0.42, 0.37{]} & {[}-0.42, 0.21{]}  & {[}-0.40, 0.20{]} \\
$\aa^{(3)}-\ba^{(3)}$   & \ {[}-0.71, 0.80{]} & {[}-0.54, 0.74{]}  & {[}-0.51, 0.72{]} \\
$\aat^{(3)}+\bat^{(3)}$ & \ {[}-0.48, 0.18{]} & {[}-0.42, 0.21{]}  & {[}-0.42, 0.20{]} \\
$\aat^{(3)}-\bat^{(3)}$ & \ {[}-0.55, 0.92{]} & {[}-0.53, 0.85{]}  & {[}-0.52, 0.85{]} \\
\midrule
$\bbb^{(+)}$            &                    & {[}-0.12, 0.18{]}  & {[}-0.11, 0.09{]} \\
$\bbb^{(-)}$            &                    & {[}-10.33, 7.14{]} & {[}-9.03, 7.16{]} \\
\bottomrule
\end{tabular}
\caption{Bounds on individual fit parameters from three separate fits of LHC, LEP and flavor data to 8 parameters (data set 1) and 14 parameters (data sets 2 and 3). The intervals correspond to $\Delta \chi^2 = 3.84$, obtained by profiling the global likelihood over all remaining parameters in the fit. The results are presented in the basis~\autoref{eq:fit-basis}.
}
\label{tab:individual-bounds}
\end{table}

Double flavor breaking along $\bbb^{(+)}$ is strongly constrained by $\Delta m_s$, whereas the orthogonal direction $\bbb^{(-)}$ is only constrained very loosely from $\sigma_{t\bar{t}b\bar{b}}$. The sensitivity in this direction can be improved with future more precise measurements of $t\bar{t}b\bar{b}$ production. It might also be indirectly improved through top-loop effects in $Z$ pole observables and $B$ meson decays in global fits, provided that the bounds on two-quark parameters that dominate those observables are strengthened.

The biggest impact of $Z$-pole observables in the global fit can be observed by comparing the results for the flavor-universal two-quark couplings from data sets 2 and 3. The bounds on $a^{(-)}$ and in particular on $a^{(3)}$ are drastically strengthened by sizeable contributions to the precision observables $\Gamma_Z$ and $R_b$, as well as correlations with $\mathcal{B}(B\to X_s\gamma)$, see \autoref{eq:all-observables}.

\paragraph{Flavor breaking in four-quark couplings} As discussed in \autoref{sec:ttbar} and \autoref{sec:dijets}, top observables probe interactions of two heavy and two light quarks, $\aa^{(w)} + \ba^{(w)}$ or $\aat^{(w)} + \bat^{(w)}$, while dijet distributions are sensitive to interactions of four light quarks, $\aa^{(w)}$ or $\aat^{(w)}$. In combination, top-antitop production and dijet angular distributions resolve all 8 directions within the subspace of flavor parameters spanned by $\{\aa^{(w)},\aat^{(w)},\ba^{(w)},\bat^{(w)}\}$.

In \autoref{fig:dijet-global}, we show bounds on two orthogonal combinations of the two four-quark couplings for weak singlets (left) and triplets (right), as we obtain them from separate fits to the three data sets introduced above.
\noindent
 To demonstrate the sensitivity of top and dijet observables to single flavor breaking, we have performed an 8-parameter fit using data set 1 only. The results (in cyan) constrain single flavor violation in four-quark couplings up to a rhomboid-shaped region. The distance between the longer parallel sides of this rhomboid corresponds to the bound on $\aa^{(w)}$ from dijet observables. The bounds on weak triplet coefficients in this direction are stronger than for weak singlets.
 The reason is that in dijet observables weak singlets and triplets probe partonic contributions with different quark flavors, see \autoref{eq:dijet-qq} and \autoref{eq:dijet-qqb}.

When adding flavor observables, single-top processes and $\sigma_{t\bar{t}b\bar{b}}$, all 14 parameters enter the observables. The results of this 14-parameter fit are shown in green. Compared to set 1, we do not observe any significant change for weak singlets (left panel). On the other hand, the combination of triplet coefficients $\aa^{(3)} + \ba^{(3)}$ is now more strongly bounded from above (right panel). This is due to contributions to single-top production, which dominate the upper bound on $(C_{qq}^{(3)})_{33ii} = \aa^{(3)} + \ba^{(3)} y_t^2$~\cite{Bruggisser:2021duo,Brivio:2019ius}.
 The other sides of the rhomboid do not change, because dijet observables dominate the bound on $\aa^{(w)}$ in the global fit.

\paragraph{Double flavor breaking} Double flavor breaking in four-quark couplings is probed in $\Delta m_s$ ($\bbb^{(+)}$) and $\sigma_{t\bar{t} b\bar{b}}$ ($\bbb^{(-)}$) at tree level. To a lesser extent, double flavor breaking also contributes to
$\mathcal{B}(B_s \to \mu^+\mu^-)$ through top loops in the same combinations $F_{qq}^{(-)}$ and $F_{qq}^{(3)}$ as loop contributions to $\Delta m_s$ do, see \autoref{eq:sbsb-loop} and \autoref{tab:sb:Bstomumu}. The partial decay rate of $Z$ bosons into bottom quarks, $\Gamma_{b\bar{b}}$, is sensitive to $\bbb^{(-)}$ and $\bbb^{(3)}$ through top loops~\cite{Dawson:2022bxd}.

In \autoref{fig:bb-bound}, we show bounds on the doubly flavor-breaking parameters obtained from our global analysis with (orange) and without (green) $Z-$pole observables.
\noindent
In the left panel, we see that the combination $\bbb^{(+)}$ is more strongly constrained than the orthogonal direction in weak gauge structures, $\bbb^{(-)}$. The reason is the extremely high sensitivity of $\Delta m_s$ to $\bbb^{(+)}$, see \autoref{eq:all-observables}, compared to the lower sensitivity of $\sigma_{t\bar{t}b\bar{b}}$ to $\bbb^{(-)}$, see \autoref{eq:ttbb-bound}. The correlation between the two parameters is due to the impact of the $\mathcal{B}(B_s \to \mu^+\mu^-)$ measurement, which slightly prefers a positive contribution to $F_{qq}^{(3)}$ and thus to $\bbb^{(3)}$~\cite{Bruggisser:2021duo}. In a fit without $\mathcal{B}(B_s \to \mu^+\mu^-)$, no such correlation would occur, as  $\Delta m_s$ and $\sigma_{t\bar{t}b\bar{b}}$ probe orthogonal directions in the parameter space.
 
Including $Z-$pole observables in the fit (orange) leads to a stronger bound on $\bbb^{(+)}$, see \autoref{fig:bb-bound}, left. At first sight, this seems surprising since $\Delta m_s$ is much more sensitive to $\bbb^{(+)}$ than $\Gamma_Z$ and $R_b$ are.
However, in our global fit the parameters $\{a^{(-)},a^{(+)},b^{(+)}\}$ are strongly constrained by $\Gamma_Z$ and $R_b$. As a consequence, cancellations between these parameters and $\bbb^{(+)}$ in $\Delta m_s$ are no longer possible. We illustrate this effect in \autoref{fig:bb-bound}, right.
The strong constraint on $a^{(-)}$ from the global fit (orange) removes a substantial part of the viable parameter space for $\bbb^{(+)}$ without $Z-$pole observables (green).
The stronger bound on $\bbb^{(+)}$ in the left panel (orange) is thus due to bounds on correlated parameters that are profiled over in the two-dimensional projection.
 
\paragraph{Flavor breaking in two-quark couplings} Similar correlation effects occur for two-quark couplings. In \autoref{fig:correlations}, left, we show flavor breaking in two-quark couplings, parametrized by $b^{(+)}$ and $b^{(-)}$.
\noindent
Adding $Z-$pole observables to the fit (orange) results in a stronger bound on $b^{(-)}$, despite the fact that $\Gamma_Z$ and $R_b$ are not sensitive to this parameter at tree level. Again, the effect is due to correlations with other parameters, as shown in \autoref{fig:correlations}, right. The constraint on $a^{(-)}$ from $Z-$pole observables (orange) removes part of the parameter space for $b^{(-)}$ that was allowed by $\mathcal{B}(B \to X_s \gamma)$, $\mathcal{B}(B_s\to \mu^+\mu^-)$, top and dijet observables (green).
These examples demonstrate the power of global fits, where correlations of SMEFT effects across different classes of observables
lead to much stronger constraints than separate fits for a sub-set of parameters.

\section{Conclusions}\label{sec:conclusions}
\noindent We have performed the first combined analysis of flavor, top, $Z-$pole and dijet observables in the SMEFT. Our main goal has been to resolve the flavor structure of SMEFT coefficients in the quark sector, assuming Minimal Flavor Violation. To this end, we have selected a set of sensitive observables that fully constrain the flavor parameter space of SMEFT operators with two and four left-handed quark doublets. Compared to our previous analysis in Ref.~\cite{Bruggisser:2021duo}, we have added $B_s$ meson mixing, $t\bar{t}b\bar{b}$ production, $Z-$pole and dijet observables to the data set. For the four-quark operators, all four sectors of observables are needed to resolve all directions in flavor space.

The observables in our analysis span three orders of magnitude in energy. We have connected SMEFT contributions at different scales through the renormalization group and reported the resulting bounds on the flavor parameters at one common energy scale in the TeV range. In this way, the results can be compared with bounds obtained from other analyses and with concrete models that could complete the SMEFT at high energies.

For the first time, we are able to fully resolve all forms of flavor breaking that such UV completions could imprint on the SMEFT coefficients. The flavor structure of four-quark operators is particularly rich, including flavor-conserving couplings, as well as flavor breaking in one or both of the quark-antiquark currents. While single flavor breaking can be probed with top-quark and/or flavor observables, double flavor breaking requires sensitivity to interactions of four third-generation quarks. We have found that $B_s-\bar{B}_s$ mixing is highly sensitive to a specific combination of doubly flavor-breaking coefficients with different weak gauge structures. The orthogonal combination can be probed in $t\bar{t}b\bar{b}$ production, but with much less sensitivity. Dijet angular distributions are pure probes of flavor-conserving SMEFT coefficients. Together with top-quark observables, they distinguish between flavor universality and single flavor breaking in four-quark couplings. 

The $Z-$pole observables, notably the total $Z$ boson decay width and the partial decay rate into bottom quarks, play an interesting role in our global analysis. Similarly to the rare meson decays $B \to X_s\gamma$ and $B_s\to \mu^+\mu^-$, these two observables are sensitive to two-quark couplings at tree level and to couplings with four heavy quarks at loop level. In the combined fit, these effects are strongly correlated between observables. This leads to indirect constraints on double flavor breaking through loop-suppressed four-quark operator contributions, which are stronger than the direct bounds from $B_s-\bar{B}_s$ mixing and $t\bar{t}b\bar{b}$ production. We stress that the potential of probing sub-leading SMEFT contributions through correlations in global fits is high. It should be further explored to resolve blind or nearly blind directions in the parameter space of Wilson coefficients.

Our results show that the flavor structure of quark interactions in the MFV-SMEFT can be fully resolved by combining sensitive observables across the energy scales. It will be interesting to conduct a similar analysis for SMEFT operators that involve right-handed quarks and/or leptons. The same strategy can also be applied to probe other flavor patterns, as predicted for instance in models with $U(2)$ flavor symmetries, with new flavor-changing neutral currents or with leptoquark couplings. Ultimately, we learn if possible new physics copies, breaks or extends the flavor symmetries of the Standard Model. If discrepancies between predictions and data are observed in the future, it will be exciting to test and compare different flavor hypotheses for their consistency with the data.

\begin{center} \textbf{Acknowledgments} \end{center}
\noindent We thank Sally Dawson and Pier Paolo Giardino for providing us with their results of four-fermion operator effects in $Z-$pole observables. The research of SB and SW has been supported by the German Research Foundation (DFG) under grant no. 396021762--TRR 257.
The work of DvD has been supported by the DFG within the Emmy Noether Programme under
grant DY-130/1-1 and the Sino-German Collaborative Research Center TRR110 ``Symmetries and the Emergence of
Structure in QCD'' (DFG Project-ID 196253076, NSFC Grant No. 12070131001, TRR 110).
%
\newpage
\appendix

\section{Flavor observables in the MFV-SMEFT}
\label{app:expressions}
In this appendix, we collect contributions of MFV-SMEFT flavor parameters to the flavor observables discussed in \autoref{sec:flavor}. The observables are linear ($\Delta m_s$) or sequilinear ($\mathcal{B}(B_s \to \mu^+\mu^-)$, $\mathcal{B}(B\to X_s \gamma$)) polynomials in terms of the flavor parameters. The contributions of the flavor parameters to $\mathcal{B}(B\to X_s \gamma)$, $\mathcal{B}(B_s \to \mu^+\mu^-)$ and $\Delta m_s$ can be found in Tabs.~\ref{tab:sb:BstoXsgamma},~\ref{tab:sb:Bstomumu} and~\ref{tab:sbsb:deltam_s}. The WET coefficients that enter the flavor observables are defined at the common scale $\mu_{\mathcal{S}} = 4.2\,$GeV. The SMEFT coefficients and the corresponding flavor parameters in the MFV-SMEFT are defined at the reference scale $\mu_0 = 2.4\,\tev$. The cutoff scale of the SMEFT is set to $\Lambda = 1\,\tev$.

\begin{table}[h]
    \resizebox{\textwidth}{!}{%
    \begin{tabular}{c | c | c c c c | c c c c c | c c c c c}
        \toprule
          & SM & $a^{(1)}$ & $b^{(1)}$ & $a^{(3)}$ & $b^{(3)}$ & $\aa^{(1)}$ & $\aat^{(1)}$ & $\ba^{(1)}$ & $\bat^{(1)}$ & $\bbb^{(1)}$ & $\aa^{(3)}$ & $\aat^{(3)}$ & $\ba^{(3)}$ & $\bat^{(3)}$ & $\bbb^{(3)}$\\ 
          \midrule
SM & 3.26 & -0.04 & -0.01 & 0.37 & -0.72 & -0.02 & 0.04 & -0.01 & 0.05 & 0.02 & 0.11 & -0.15 & 0.01 & -0.25 & -0.20\\ 
 \midrule
$a^{(1)}$ &  & 0 & 0 & 0 & 0 & 0 & 0 & 0 & 0 & 0 & 0 & 0 & 0 & 0 & 0\\ 
$b^{(1)}$ &  &  & 0 & 0 & 0 & 0 & 0 & 0 & 0 & 0 & 0 & 0 & 0 & 0 & 0\\ 
$a^{(3)}$ &  &  &  & 0.01 & -0.04 & 0 & 0 & 0 & 0 & 0 & 0.01 & -0.01 & 0 & -0.01 & -0.01\\ 
$b^{(3)}$ &  &  &  &  & 0.04 & 0 & 0 & 0 & -0.01 & 0 & -0.01 & 0.02 & 0 & 0.03 & 0.02\\ 
 \midrule
$\aa^{(1)}$ &  &  &  &  &  & 0 & 0 & 0 & 0 & 0 & 0 & 0 & 0 & 0 & 0\\ 
$\aat^{(1)}$ &  &  &  &  &  &  & 0 & 0 & 0 & 0 & 0 & 0 & 0 & 0 & 0\\ 
$\ba^{(1)}$ &  &  &  &  &  &  &  & 0 & 0 & 0 & 0 & 0 & 0 & 0 & 0\\ 
$\bat^{(1)}$ &  &  &  &  &  &  &  &  & 0 & 0 & 0 & 0 & 0 & 0 & 0\\ 
$\bbb^{(1)}$ &  &  &  &  &  &  &  &  &  & 0 & 0 & 0 & 0 & 0 & 0\\ 
$\aa^{(3)}$ &  &  &  &  &  &  &  &  &  &  & 0 & 0 & 0 & 0 & 0\\ 
$\aat^{(3)}$ &  &  &  &  &  &  &  &  &  &  &  & 0 & 0 & 0.01 & 0\\ 
$\ba^{(3)}$ &  &  &  &  &  &  &  &  &  &  &  &  & 0 & 0 & 0\\ 
$\bat^{(3)}$ &  &  &  &  &  &  &  &  &  &  &  &  &  & 0 & 0.01\\ 
$\bbb^{(3)}$ &  &  &  &  &  &  &  &  &  &  &  &  &  &  & 0 \\
        \bottomrule
    \end{tabular}
    }
\caption{Contributions of flavor coefficients to $\mathcal{B}(\bar{B}_s\to X_s\gamma)\times 10^4$. The reference scale is set to $\mu_0 = 2.4\,\tev$; the cutoff scale is fixed to $\Lambda = 1\,$TeV. The contributions are rounded to two decimal places.\label{tab:sb:BstoXsgamma}}
\end{table}

\begin{table}[h]
    \resizebox{\textwidth}{!}{%
    \begin{tabular}{c | c | c c c c | c c c c c | c c c c c}
        \toprule
         & SM & $a^{(1)}$ & $b^{(1)}$ & $a^{(3)}$ & $b^{(3)}$ & $\aa^{(1)}$ & $\aat^{(1)}$ & $\ba^{(1)}$ & $\bat^{(1)}$ & $\bbb^{(1)}$ & $\aa^{(3)}$ & $\aat^{(3)}$ & $\ba^{(3)}$ & $\bat^{(3)}$ & $\bbb^{(3)}$\\ 
         \midrule
SM & 3.57 & 0.32 & -36.91 & -3.02 & -42.45 & -0.07 & 3.06 & 2.84 & 5.94 & 5.75 & 1.30 & -3.35 & -0.79 & -5.39 & -4.11\\
\midrule
$a^{(1)}$ &  & 0.01 & -1.64 & -0.13 & -1.88 & 0 & 0.14 & 0.13 & 0.26 & 0.25 & 0.06 & -0.15 & -0.03 & -0.24 & -0.18\\ 
$b^{(1)}$ &  &  & 95.51 & 15.61 & 219.65 & 0.34 & -15.86 & -14.71 & -30.73 & -29.76 & -6.75 & 17.34 & 4.07 & 27.88 & 21.25\\ 
$a^{(3)}$ &  &  &  & 0.64 & 17.95 & 0.03 & -1.30 & -1.20 & -2.51 & -2.43 & -0.55 & 1.42 & 0.33 & 2.28 & 1.74\\ 
$b^{(3)}$ &  &  &  &  & 126.29 & 0.39 & -18.24 & -16.92 & -35.33 & -34.22 & -7.76 & 19.94 & 4.68 & 32.06 & 24.43\\
\midrule
$\aa^{(1)}$ &  &  &  &  &  & 0 & -0.03 & -0.03 & -0.06 & -0.05 & -0.01 & 0.03 & 0.01 & 0.05 & 0.04\\ 
$\aat^{(1)}$ &  &  &  &  &  &  & 0.66 & 1.22 & 2.55 & 2.47 & 0.56 & -1.44 & -0.34 & -2.32 & -1.76\\ 
$\ba^{(1)}$ &  &  &  &  &  &  &  & 0.57 & 2.37 & 2.29 & 0.52 & -1.34 & -0.31 & -2.15 & -1.64\\ 
$\bat^{(1)}$ &  &  &  &  &  &  &  &  & 2.47 & 4.79 & 1.09 & -2.79 & -0.66 & -4.49 & -3.42\\
$\bbb^{(1)}$ &  &  &  &  &  &  &  &  &  & 2.32 & 1.05 & -2.70 & -0.63 & -4.34 & -3.31\\ 
$\aa^{(3)}$ &  &  &  &  &  &  &  &  &  &  & 0.12 & -0.61 & -0.14 & -0.99 & -0.75\\ 
$\aat^{(3)}$ &  &  &  &  &  &  &  &  &  &  &  & 0.79 & 0.37 & 2.53 & 1.93\\ 
$\ba^{(3)}$ &  &  &  &  &  &  &  &  &  &  &  &  & 0.04 & 0.59 & 0.45\\
$\bat^{(3)}$ &  &  &  &  &  &  &  &  &  &  &  &  &  & 2.04 & 3.10\\ 
$\bbb^{(3)}$ &  &  &  &  &  &  &  &  &  &  &  &  &  &  & 1.18 \\
        \bottomrule
    \end{tabular}
    }
    \caption{Contributions of flavor coefficients to $\mathcal{B}(\bar{B}_s\to \mu^+\mu^-)\times 10^9$. The reference scale is set to $\mu_0 = 2.4\,\tev$; the cutoff scale is fixed to $\Lambda = 1\,$TeV. The contributions are rounded to two decimal places.}\label{tab:sb:Bstomumu}
\end{table}

\begin{table}[t]
    \resizebox{\textwidth}{!}{%
    \begin{tabular}{c | c c | c c | c c c c c | c c c c c}
        \toprule
        SM & $a^{(1)}$ & $b^{(1)}$ & $a^{(3)}$ & $b^{(3)}$ & $\aa^{(1)}$ & $\aat^{(1)}$ & $\ba^{(1)}$ & $\bat^{(1)}$ & $\bbb^{(1)}$ & $\aa^{(3)}$ & $\aat^{(3)}$ &  $\ba^{(3)}$ & $\bat^{(3)}$ & $\bbb^{(3)}$ \\
        \midrule
        17.28 & 0.28 & -3.25 & -4.60 & 5.97 & 0.10 &  2.09 & -3.16 & -1.27 & 133.02 & 4.01 & -1.82 & -0.53 & -4.51 & 132.91 \\
        \bottomrule
    \end{tabular}
    }
    \caption{Contributions of flavor coefficients to $\Delta m_s$ in units of $\text{ps}^{-1}$. The reference scale is set to $\mu_0 = 2.4\,\tev$; the cutoff scale is fixed to $\Lambda = 1\,$TeV. The contributions are rounded to two decimal places.}\label{tab:sbsb:deltam_s}
\end{table}


\newpage
\bibliographystyle{JHEP}
\bibliography{flavor-in-smeft}


\end{document}